\documentclass[nofootinbib,aps,prd,groupedaddress,preprintnumbers,%
  showpacs,showkeys,floatfix,amssymb,amsfonts]{revtex4-1}  
\usepackage{hyperref}
\usepackage[usenames]{color}
\usepackage{todonotes}
\usepackage{bbold}
\usepackage{amsmath}
\usepackage{multirow}
\usepackage{graphicx}
\usepackage{xfrac}
\usepackage[utf8]{inputenc}
\newcommand{\tr}{\operatorname{Tr}} 
\usepackage{amsfonts}
\usepackage{amssymb}
\usepackage{extarrows}

\usepackage[sort&compress]{natbib}

\newcommand{\be}{\begin{equation}}
\newcommand{\ee}{\end{equation}}
\newcommand{\bea}{\begin{eqnarray}}
\newcommand{\eea}{\end{eqnarray}}

\newcommand{\MSb}{\overline{\text{MS}}}

\newcommand{\numax}{\nu_{\textrm{\rm max}}}
\newcommand{\zmax}{z_{\textrm{\rm max}}}

\let\oldbibliography\thebibliography
\renewcommand{\thebibliography}[1]{\oldbibliography{#1}
\setlength{\baselineskip}{9.5pt}
\setlength{\itemsep}{5pt}} 

\begin{document}

\title{Gluon PDF of the proton using twisted mass fermions}

\author{
  Joseph Delmar$^{1}$,
  Constantia Alexandrou$^{2,3}$,
 Krzysztof Cichy$^4$,
  Martha Constantinou$^{1}$,
Kyriakos Hadjiyiannakou$^{2,3}$ 
}

\affiliation{
  \vskip 1 cm
    $^1$Department of Physics, Temple University, 1925 N. 12th Street, Philadelphia, PA 19122-1801, USA\\ 
  \vskip 0.05cm
  $^2$Department of Physics, University of Cyprus,  P.O. Box 20537,  1678 Nicosia, Cyprus\\
  \vskip 0.05cm
   $^3$Computation-based Science and Technology Research Center,
  The Cyprus Institute, 20 Kavafi Str., Nicosia 2121, Cyprus \\
 \vskip 0.05cm
  $^4$ Faculty of Physics, Adam Mickiewicz University, Uniwersytetu Pozna\'{n}skiego 2, 61-614 Pozna\'{n}, Poland \\
  \vskip 0.05cm
  }

\begin{abstract}

In this paper, we present lattice QCD results for the $x$-dependence of the unpolarized gluon PDF for the proton. 
We use one ensemble of $N_f=2+1+1$ maximally twisted mass fermions with a clover improvement, and the Iwasaki improved gluon action. 
The quark masses are tuned to produce a pion with a mass of 260 MeV. 
The ensemble has a lattice spacing of $a=0.093$ fm and a spatial extent of 3 fm. 
We employ the pseudo-distribution approach, which relies on matrix elements of non-local operators that couple to momentum-boosted hadrons. 
In this work, we use five values of the momentum boost between 0 and 1.67 GeV. 
The gluon field strength tensors of the non-local operator are connected with straight Wilson lines of varying length $z$. 
The light-cone Ioffe time distribution (ITD) is extracted utilizing data with $z$ up to 0.56 fm and a quadratic parametrization in terms of the Ioffe time at fixed values of $z$. 
We explore systematic effects, such as the effect of the stout smearing for the gluon operator, excited states effects, and the dependence on the maximum value of $z$ entering the fits to obtain the gluon PDF. 
Also, for the first time, the mixing with the quark singlet PDFs is eliminated using matrix elements with non-local quark operators that were previously analyzed within the quasi-PDF framework on the same ensemble. 
Here, we expand the data set for the quark singlet and reanalyze within the pseudo-PDFs method eliminating the corresponding mixing in the gluon PDF.

\end{abstract}


\maketitle
	
\section{Introduction}
\label{sec:intro}

As the mediators of the strong force, gluons play a significant role in the internal structure of hadrons. However, color confinement, a key aspect of quantum chromodynamics (QCD), prevents direct observation of quarks and gluons. Instead, both theoretical and experimental approaches to hadronic structure calculations rely on QCD factorization, which separates the perturbatively-calculable hard-scattering part from the non-perturbative part described by form factors and distribution functions, including parton distribution functions (PDFs). PDFs are probability distributions quantifying the likelihood of finding partons with a particular momentum fraction. Precise and accurate calculations of the gluon PDF are necessary for $J/\psi$ photo production at Jefferson lab, the cross-section of Higgs boson production and jet production at the Large Hadron Collider (LHC), as well as providing theoretical input to experiments at the future Electron-Ion Collider (EIC) in the U.S. and the Electron-Ion Collider in China (EicC).

Lattice QCD is a first-principles approach to calculating strong force quantities performed on a discrete 4-dimensional Euclidean lattice. 
While lattice QCD calculations have proven successful in extracting the non-perturbative dynamics of QCD governing hadron structure, the light-like nature of PDFs prevents direct calculation on Euclidean lattices. Several methods have been proposed over the last decade to relate lattice data to physical light-cone distributions. Two notable and most widely used approaches are the quasi-distribution~\cite{PhysRevLett.110.262002, Ji:2014sciC} and pseudo-distribution~\cite{RADYUSHKIN2017314, PhysRevD.96.034025, RADYUSHKIN2018433, PhysRevD.98.014019, RADYUSHKINintJModPhys} methods. These approaches utilize the same matrix elements of momentum-boosted hadrons coupled to non-local operators containing a Wilson line but differ in the way the Euclidean observable is factorized into its light-cone counterpart directly in coordinate space (pseudo) or after reconstruction of the $x$-dependence, i.e.\ in momentum space (quasi).
Typically, they are also renormalized differently.
By construction, the renormalization for pseudo-distributions employs canceling the divergences by forming an appropriate ratio of matrix elements (ratio scheme).
In turn, quasi-distributions are typically renormalized using a dedicated calculation of vertex functions of the operator under study that leads to an RI/MOM type of renormalization.
It should be noted that, the ratio scheme is also increasingly utilized for quasi-distributions in hybrid schemes~\cite{Ji:2020brr} that treat short and long scales differently. 
Another typical difference is in the $x$-dependence reconstruction.
For quasi-distributions, this step uses Euclidean matrix elements in the full range of the non-local operator lengths, $z$.
Pseudo-distributions, in turn, are matched in coordinate space, which imposes limitations on the value of $z$, which needs to be kept relatively small so that it remains in the perturbative region.
Thus, without access to the full range of $z$, approaches based on pseudo-distributions typically employ a physically-motivated fitting ansatz for the functional form of the reconstructed function.

There have been several lattice calculations of various types of quark distributions for the nucleon and other hadrons (mostly the pion), see e.g.\ Refs.~\cite{Lin:2014zya,Alexandrou:2015rja,Chen:2016utp,Alexandrou:2016jqi,Zhang:2017bzy,Alexandrou:2017huk,Zhang:2017zfe,Alexandrou:2018pbm,Alexandrou:2018eet,Liu:2018uuj,Zhang:2018nsy,Alexandrou:2019lfo,Izubuchi:2019lyk,Cichy:2019ebf,Chai:2020nxw,Zhang:2020gaj,Bhattacharya:2020xlt,Bhattacharya:2020jfj,Alexandrou:2020zbe,Alexandrou:2020uyt,Alexandrou:2020qtt,Lin:2020ssv,Fan:2020nzz,Gao:2020ito,Bringewatt:2020ixn,Hua:2020gnw,Alexandrou:2021oih,Alexandrou:2021bbo, Bhattacharya:2021moj,Gao:2021dbh,Hua:2022kcm,Gao:2022iex,Bhattacharya:2022aob,LatticeParton:2022xsd,Bhattacharya:2023nmv} for quasi-distributions, Refs.~\cite{Orginos:2017kos,Karpie:2018zaz,Karpie:2019eiq,Joo:2019jct,Joo:2019bzr,Joo:2020spy,Bhat:2020ktg,DelDebbio:2020rgv,Karpie:2021pap,Egerer:2021ymv,HadStruc:2021qdf,Bhat:2022zrw,HadStruc:2022nay} for pseudo-distributions and
Refs. ~\cite{Cichy:2018mum,Ji:2020ect,Constantinou:2020pek,Cichy:2021lih,Cichy:2021ewm} for recent reviews.
However, the gluonic component of hadron structure has been less studied, though the contribution to various physical quantities is significant. Phenomenological data and lattice calculations, for instance, suggest that gluons account for approximately $40\%$ of the hadron's momentum at a scale of $6.25 \, \text{GeV}^2$ \cite{PhysRevD.89.054028, PhysRevD.96.054503}. It is essential to better understand how the gluon contributes to hadron structure, which has led to several dedicated lattice calculations~\cite{Fan:2018dxu,Fan:2020cpa,HadStruc:2021wmh,HadStruc:2022yaw} and phenomenological analyses of experimental data sets~\cite{NNPDF:2017mvq,Hou:2019efy,Moffat:2021dji} on this topic. As has been done in the case of quark PDFs~\cite{PhysRevD.104.016015,NNPDF,PhysRevD.93.114017}, lattice data on $x$-dependent quantities have the potential to assist in constraining global analyses.

In this work, we present our calculation of the unpolarized gluon PDF for the proton using the pseudo-PDF approach. We calculate the Ioffe-time pseudo-distribution function (pseudo-ITD) by taking the ratio of matrix elements and evolving to a common scale. The ITD describes the interaction of the nucleon with the probe in deep inelastic scattering (DIS) interactions. We use a fitting ansatz to reconstruct the pseudo-PDF from the pseudo-ITD. This approach has proved successful for the extraction of the quark pseudo-PDF. The gluon component presents additional difficulties, including the need for an order of magnitude more statistics arising from the noise associated with the purely disconnected diagram. The gluon PDF also mixes with the quark singlet PDF. Previous lattice calculations have neglected this mixing. We present the first analysis incorporating the quark singlet mixing from lattice QCD data. We compare our pseudo-PDF results neglecting mixing with lattice results from the HadStruc collaboration~\cite{HadStruc:2021wmh}. We also compare our results with and without mixing to global analysis from the JAM collaboration~\cite{PhysRevD.104.016015}.

This paper is organized as follows. In Secs.~\ref{sec:setup} and \ref{sec:latt_details}, we describe the theoretical and lattice setups for the calculation. In Sec.~\ref{sec:MEs_and_DRs}, we present our analysis of various smearing and source-sink time separation values of the matrix elements and reduced-ITDs. Sec.~\ref{sec:no_mixing} shows the results of the pseudo-ITD and pseudo-PDF neglecting mixing with the quark singlet, and Sec.~\ref{sec:mixing} presents the results addressing the mixing with the quark singlet.

\section{Methodology}

\subsection{Approach}
\label{sec:setup}

The computationally expensive component of the methodology is the evaluation of matrix elements with momentum-boosted proton states, $N(P)$, that couple to non-local gluon operators; $P$ indicates the proton momentum. 
The operator is constructed by two gluon field-strength tensors, $F^{\mu\nu}$, located at two lattice points that are spatially separated in the $\hat{z}$ direction by distance $z$. 
The operator also contains two straight Wilson lines, connecting points $0\to z$ and $z\to0$, to ensure gauge invariance. 
The matrix element reads
\begin{equation}
    \label{eq:gluon_oper}
    M_{\mu i; \nu j}(P,z)  = \langle N(P)| F_{\mu i}(z) W(z, 0) F_{\nu j}(0)  W(0,z)|N(P) \rangle \,,
\end{equation}
where $F_{\mu\nu}$ is the gluon field strength tensor defined as
{\small{
\begin{eqnarray}
    F_{\mu\nu} (x) &=& \frac{i}{8 g_0} \bigg[ U_\mu(x) U_\nu(x+a\hat{\mu}) U^\dag_\mu(x+a \hat{\nu}) U^\dag_\nu(x) + U_\nu(x) U^\dag_\mu(x+a\hat{\nu}-a\hat{\mu}) U_\nu^\dag(x-a\hat{\mu}) U_\mu(x-a\hat{\mu}) \nonumber \\
    && \qquad + U^\dag_\mu(x-\hat{\mu}) U^\dag_\nu(x-a\hat{\nu}-a\hat{\mu}) U_\mu(x-a\hat{\nu}-a\hat{\mu}) U_\nu(x-a\hat{\nu}) \nonumber \\
    && \qquad+ U^\dag_\nu(x-a\hat{\nu}) U_\mu(x-a\hat{\nu}) U_\nu(x-a\hat{\nu}+a\hat{\mu}) U^\dag_\mu(x) - h.c \bigg]\,,
    \label{Eq:FST}
\end{eqnarray}
}}
and $g$ is the bare coupling constant. 
Potential candidates for the gluon operator are given below for different values of the indices $\mu\,,\nu,\, i,\, j$, which can be temporal or spatial, that is
\begin{eqnarray}
{\cal O}_0 &\equiv& \sum_{i<j} F_{ij}(x+z \hat{z}) W(x+z \hat{z},x) F_{ij}(x) W(x,x+z \hat{z}) \nonumber \\ &-& \sum_i F_{it}(x+z \hat{z}) W(x+z \hat{z},x) F_{it}(x)W(x,x+z \hat{z})\,, \\[0.5ex]
{\cal O}_1 &\equiv& \frac{1}{2} \sum_{i} F_{it}(x+z \hat{z}) W(x+z \hat{z},x) F_{it}(x) W(x,x+z \hat{z})\,, \quad i \ne z\,,  \\[0.5ex]
{\cal O}_2 &\equiv& \frac{1}{2} \sum_{i} F_{iz}(x+z \hat{z}) W(x+z \hat{z},x) F_{iz}(x) W(x,x+z \hat{z})\,,  \\[0.5ex]
{\cal O}_3 &\equiv& \frac{1}{2} \sum_{i} F_{it}(x+z \hat{z}) W(x+z \hat{z},x) F_{iz}(x) W(x,x+z \hat{z})\,, \quad i \ne z\,,  \\[0.5ex]
{\cal O}_4 &\equiv& \frac{1}{2} \sum_i F_{it}(x+n\hat{k}) W(x+n\hat{k},x) F_{it}(x) \nonumber \\ 
&-& \sum_{ij} F_{ij}(x+n\hat{k}) W(x+n\hat{k},x) F_{ij}(x)\,, \quad i \ne j \ne z \,. 
\end{eqnarray}

The various options of the indices lead to the construction of operators with different properties. 
Here, we use the operator ${\cal O}_4$, which does not exhibit mixing under renormalization. 
This operator has a non-vanishing vacuum expectation value that must be subtracted. 
Since the calculation of the gluon loops is computationally very inexpensive, the vacuum expectation value subtraction does not pose any challenges in the calculation. 
It should be noted that, regardless of the choice of operator, the unpolarized gluon PDF mixes with the unpolarized singlet quark PDF. 
We take this mixing into account in our analysis and we quantify its effects by comparing to results with the mixing neglected.

The matrix elements of Eq.~\eqref{eq:gluon_oper}, $M_{\mu i; \nu j}$, are extracted from the ground state contribution to the ratio 
\begin{equation}
 R_{\cal O}(t_s,\tau,t_0;P,z) = \frac{C^{\text{3pt}}_{\cal O}(t_s,\tau; P, z)}{C^{\text{2pt}} (t_s; P)}\,\, \stackrel{t_s<\tau<t'} \longrightarrow \,\, \frac{4}{3} \left( \frac{m^2}{4 E} - E \right) M_{\cal O}(t_s; P,z)\,,
  \label{Eq:Ratio}
\end{equation}
taken between the three-point and two-point correlation functions.
The variables $t_s$, $\tau$, and $t_0$ indicate the time of the sink,
operator insertion, and source, respectively.
Without loss of generality, we have taken the source position to be at $t_0=0$. 
The ground state contribution, $M_{\cal O}\equiv M_{\mu i; \nu j}$, is identified at large enough values of $t_s$ and at $\tau$ away from the source and the sink.
Practically, we seek convergence with a variance of $t_s$ and at $\tau$.

For the calculation of gluonic contributions to the proton, $C^{\text{3pt}}_{\cal O}$ correspond to the so-called disconnected contributions, which are constructed by the expectation value of a product of a gluon loop with the proton two-point function. 
Also, for the unpolarized gluon PDF, the appropriate parity projector is $\Gamma_0 \equiv \frac{1}{4} (1+\gamma_0)$ for both the three- and two-point functions.

In our analysis, we implement the pseudo-ITD framework, which requires several nontrivial steps to extract the $x$-dependence of the gluon PDF. For convenience, we use $M_{g}$ to denote the ground state contribution for the operator ${\cal O}_4$. First, the matrix elements at different values of $P$ and $z$ are combined to construct the reduced Ioffe-time distribution (pseudo-ITD),
\begin{equation}
    \label{eq:double_ratio}
    \mathfrak{M}_g(\nu,z^2) \equiv \bigg( \frac{M_g(\nu,z^2)}{M_g(\nu,0)|_{z=0}} \bigg) \bigg/ \bigg( \frac{M_g(0,z^2)|_{p=0}}{M_g(0,0)|_{p=0,z=0}} \bigg)\,,
\end{equation}
which depends on the Lorentz-invariant quantities $\nu\equiv z\cdot P$ (Ioffe time) and $z^2$.
For multiplicatively renormalizable operators, the reduced ITD acts as a gauge invariant renormalization scheme that removes UV divergences, including the power divergence due to the presence of the Wilson line. 
The effects of the residual scale $1/z$ can be accounted for by an evolution term (see below) and data from different scales $1/z$ can be combined into ITDs defined at a common renormalization scale, $\mu^2$. 
Furthermore, it is anticipated that Eq.~\eqref{eq:double_ratio} leads to suppressed discretization and higher-twist effects, which are assumed similar in the two single ratios shown above~\cite{Orginos:2017kos}. 

\medskip
Another component of this work is the calculation of the unpolarized quark PDF to address the mixing with the gluon case.
The matrix element can be written similarly to the gluon case, that is
\begin{equation}
\label{eq:matrix_el}
    M_f(z,P) = \langle N(P)|\overline{\psi}_f(z)\, \gamma^0 \, W(z) \psi_f(0)|N(P)\rangle\,,
\end{equation}
where the fermionic field $\psi_f(x) \equiv \psi_f(\vec{x},t)$ is taken to be the up, down, and strange quark; $f$ indicates the flavor.
For a proper flavor decomposition of the up and down quark contributions, we calculate the disconnected diagram in addition to the connected one.
Moreover, the strange-quark contribution is purely disconnected for the nucleon case.
Forming the quark-disconnected contributions requires the evaluation of quark loops that are combined with the nucleon two-point correlators. The quark loop of the non-local operator reads
\begin{equation}
\label{eq:loop}
    \mathcal{L}(t_{\rm ins},z) =
\sum\limits_{\vec{x}_{\rm ins}} \tr\left[D_q^{-1}(x_{\rm ins};x_{\rm ins}+z)\gamma^0 W(x_{\rm ins},x_{\rm ins}+z)\right]\,,
\end{equation}
where $D_f^{-1}(x_{\rm ins};x_{\rm ins}+z)$ is the quark propagator, whose endpoints are connected by a Wilson line. 
More details in the calculation of the disconnected contributions can be found in Ref.~\cite{Alexandrou:2021oih}.
Here we combine the connected and disconnected contributions to the matrix element to form the singlet $u+d+s$ combination, $M_{q}$.
The latter is normalized by constructing the pseudo-ITD, $\mathfrak{M}_{q}$ similarly to the definition of Eq.~\eqref{eq:double_ratio}.

\medskip
To extract the light-cone counterpart of $\mathfrak{M}_g$, indicated as $Q_{gq}$, one must apply a matching procedure known to one-loop-level accuracy~\cite{Balitsky:2019krf,Balitsky:2021bds},
\begin{eqnarray}
    \label{eq:itd_equation}
    Q_{gq}(\nu,z^2,\mu^2) = \mathfrak{M}_g(\nu,z^2)\,\langle x\rangle_g^{\mu} &+& \frac{\alpha_s N_c}{2\pi}\int_0^1 du \; \mathfrak{M}_g(u\nu,z^2)\,\langle x\rangle_g^{\mu} \bigg\{ \text{ln}\bigg(\frac{z^2 \mu^2 \text{e}^{2\gamma_E}}{4}\bigg)\mathfrak{B}_{gg}(u) + L(u) \bigg\} \nonumber \\
    &+&  \frac{\alpha_s C_F}{2 \pi} \, \mathrm{ln}\bigg( \frac{z^2 \mu^2 e^{2 \gamma_E}}{4} \bigg) \int_0^1 du \;  \left(\mathfrak{M}_S (u \nu, \mu^2)-\mathfrak{M}_S (0, \mu^2)\right) \; \mathfrak{B}_{gq}(u) \,,
\end{eqnarray}
where $\langle x\rangle_g^{\mu}$ is the gluon momentum fraction renormalized at the scale $\mu$,  and  
\begin{equation}
\mathfrak{M}_S(\nu, \mu^2)=\int_0^1 dx \sum_f \cos(x\nu) x \left(q_f(x, \mu^2)+\bar{q}_f(x, \mu^2)\right)\,, 
\end{equation}
with $q_f(x, \mu^2)$ ($\bar q_f(x, \mu^2)$) being the quark (antiquark) PDF of flavor $f$,
and the sum runs over all considered quark flavors ($f=u,d,s$).
This distribution is related to the imaginary part of the double ratio $\mathfrak{M}_q$,
\begin{equation}
\textrm{Im}\, \mathfrak{M}_q(\nu, \mu^2) = \int_0^\nu dy\, \mathfrak{M}_S(y, \mu^2)\,.
\end{equation}
Differentiating this equation with respect to the upper limit of the integral, we get
\begin{equation}
\mathfrak{M}_S(\nu, \mu^2)=\frac{d\textrm{Im}\, \mathfrak{M}_q(\nu, \mu^2)}{d\nu}\,.    
\end{equation}
Thus, the singlet quark Ioffe-time distribution appearing in the matching equation, $\mathfrak{M}_S (\nu, \mu^2)$, is purely real and related to the imaginary part of the quark double ratio.

The matching kernels read
\begin{equation}    \label{eq:B_Lkernel}
    \mathfrak{B}_{gg}(u) = 2\bigg[\frac{(1-u(1-u))^2}{1-u}\bigg]_+\,,\qquad 
    L(u)= 4 \bigg[\frac{u+\text{ln}(1-u)}{1-u}\bigg]_{+} + \frac{2}{3} \big[1-u^3\big]_{+}\,,\qquad  \mathfrak{B}_{gq}(u) =  1 + (1-u)^2 \,,
\end{equation}
and the plus prescription is given by
$\int^1_0 [f(u)]_+ \mathfrak{M}_g(u\nu)=\int^1_0 f(u) (\mathfrak{M}_g(u\nu)-\mathfrak{M}_g(\nu))$.

The matching equations involve evolving the reduced gluon ITD to a common scale ($\mathfrak{B}_{gg}(u)$ term), converting the expressions to the light-cone gluon ITD in the $\rm \overline{MS}$ scheme ($L(u)$ term) and taking its mixing with the singlet quarks into account ($\mathfrak{B}_{gq}$ term). 
It is convenient to rewrite Eq.~\eqref{eq:itd_equation} in three parts so that one can inspect the role of the three terms separately,
\begin{eqnarray}
\label{eq:evol}
\mathfrak{M}_g'(\nu,z^2,\mu^2) = \mathfrak{M}_g(\nu,z^2) + \frac{\alpha_s N_c}{2\pi}\int_0^1 du \; \mathfrak{M}_g(u\nu,z^2) \,\text{ln}\bigg(\frac{z^2 \mu^2 \text{e}^{2\gamma_E}}{4}\bigg)\mathfrak{B}_{gg}(u)\,,
\end{eqnarray}
where $\mathfrak{M}_g'(\nu,z^2,\mu^2)$ is the evolved gluon ITD, which depends on $\nu$, the final scale $\mu^2$ and the initial scale $z^2$. 
The matching and conversion to the $\overline{\rm MS}$ scheme is given by
\begin{equation}
\label{eq:match}
Q_g(\nu,z^2,\mu^2) = \mathfrak{M}_g'(\nu,z^2,\mu^2) + \frac{\alpha_s N_c}{2\pi}\int_0^1 du \; \mathfrak{M}_g(u\nu,z^2)\, L(u)\,.
\end{equation}
Finally, we take the mixing with the singlet quark into account, $\mathfrak{M}_q$, arriving at the final light-cone ITD,
\begin{equation}
\label{eq:mix}
Q_{gq}(\nu,z^2,\mu^2) = Q_g(\nu,z^2,\mu^2)\,\langle x\rangle_g^{\mu} + \frac{\alpha_s C_F}{2 \pi} \, \mathrm{ln}\bigg( \frac{z^2 \mu^2 e^{2 \gamma_E}}{4} \bigg) \int_0^1 du \;  \left(\mathfrak{M}_S (u \nu, \mu^2)-\mathfrak{M}_S (0, \mu^2)\right) \; \mathfrak{B}_{gq}(u) \,.\end{equation}
The matched gluon ITD still keeps track of the initial scale $z^2$ at this stage.
However, different scales $z^2$ should lead to the same light-cone ITDs up to higher-twist effects.
For data points where this holds, i.e.\ leading to consistent values of $Q_g(\nu,z^2,\mu^2)$ from different initial $z^2$, $Q_g$ is averaged over the same values of $\nu$ extracted from different combinations of $P$ and $z$. We denote such $\nu$-averaged ITDs by $Q_{g/gq}(\nu,\mu^2)$, i.e.\ dropping the argument indicating the initial scale $z$.

To extract the $x$-dependent gluon PDF, $xg(x)$, the light-cone ITDs need to be subjected to a cosine Fourier transform, 
\begin{equation}
\label{eq:PDF2ITD}
Q_{g/gq}(\nu,\mu^2) =\int_{0}^1 dx \, \cos(\nu x) x g(x,\mu^2)\,.
\end{equation}
The extraction of $xg(x,\mu^2)$ poses an inverse problem~\cite{Karpie:2018zaz}, because one attempts to calculate a continuous distribution from a limited number of lattice data points for a finite range of Ioffe times up to some $\numax$. 
Therefore, to determine $xg(x,\mu^2)$, one requires additional information, which can be chosen in several ways. 
Here, we reconstruct the gluon PDF by using a fitting ansatz commonly used in the analysis of experimental data sets, that is
\begin{equation}
\label{eq:ansatz}
x q(x) = N x^a (1-x)^b,
\end{equation}
where the exponents $a,\,b$ are fitting parameters and $N$ is the normalization that is fixed by the gluon momentum fraction $\int_0^1 dx \,x g(x) = \langle x \rangle_g$.
The lattice data are, thus, fitted according to the minimization of
\begin{equation}
    \label{eq:chi_sq}
    \chi^2 = \sum_{\nu=0}^{\nu_{\rm max}} \frac{\big(Q_{g/gq}(\nu,\mu^2)-Q_f(\nu, \mu^2)\big)^2}{\sigma^2_Q(\nu, \mu^2)}\,.
\end{equation}
We consider the reconstruction in the cases with ($Q_{gq}$) and without ($Q_g$) the mixing taken into account, to assess the effect of this mixing at the level of the $x$-dependent distributions.
The data are weighted by the inverse variance of the light-cone ITDs, $\sigma_{Q_{g/gq}(\nu,\mu^2)}^2$.
$Q_f(\nu, \mu^2)$ is the cosine Fourier transform of the assumed fitting ansatz.


\subsection{Setup of lattice calculation}
\label{sec:latt_details}
\vspace*{-2mm}

The calculation is performed using an $N_f=2+1+1$ ensemble of twisted-mass clover-improved fermions and Iwasaki-improved gluons~\cite{Alexandrou:2018egz}. 
The quark masses are fixed such that the pion has approximately twice its physical mass ($m_\pi = 260$ MeV). 
The lattice spacing is $a=0.0938(2)(3)$ fm, and the lattice volume is $32^3 \times 64$ in lattice units. 
The parameters of the ensemble are summarized in Table~\ref{tab:params}. 

\begin{table}[h!]
\begin{center}
\renewcommand{\arraystretch}{1.8}
\begin{tabular}{l| c  c  c  c  c  c c }
    \hline
    Ensemble   & $\qquad\beta\qquad$ & $\qquad a$ [fm] $\qquad$ & $\quad$volume $L^3\times T$ $\quad$& $\quad$$N_f$ $\quad$&$\quad$ $m_\pi$ [MeV] $\quad$&
 $\quad$   $L m_\pi$$\quad$ &$\quad$ $L$ [fm]$\quad$\\
    \hline
    cA211.30.32 & 1.726 & 0.0938(2)(3)  & $32^3\times 64$  & 2+1+1 & 260
    & 4 & 3.0 \\
    \hline
\end{tabular}
\caption{\small{Parameters of the ensemble used in this work.}}
\label{tab:params}
\end{center}
\end{table}

Matrix elements of gluon operators have increased gauge noise, and one needs to (a) obtain high statistics and (b) use smoothing techniques. 
To this end, we calculate the correlation functions from different source positions on the same configuration, as the cA211.30.32 ensemble has about 1,200 thermalized gauge configurations~\cite{Alexandrou:2018egz}. 
Utilizing several source positions per configuration combined with the large speed-up achieved with the use of the multi-grid~\cite{Clark:2016rdz,Alexandrou:2016izb,Bacchio:2017pcp,Alexandrou:2018wiv}, leads to an efficient increase in statistics. Here,
we analyze a total of 200 source positions for each configuration. 
To further increase statistics without loss of generality, we calculate the matrix element using six kinematically equivalent setups, where both the Wilson line and momentum boost are in the $\pm x,\,\pm y,\,\pm z$ directions. 
These six matrix elements can be averaged over, leading to total statistics exceeding one million measurements, as shown in Table~\ref{tab:stat}. 
Since the pseudo-ITD utilizes matrix elements at several values of the proton momentum, we use five values, that is, $P=0,\,0.42,\, 0.83,\, 1.25,\, 1.67$ GeV. 
Each matrix element is normalized with the $P=0$ case and we found non-negligible correlations between the numerator and denominator of the reduced ITD in Eq.~\eqref{eq:double_ratio}.
These are eliminated by calculating all matrix elements at the same configurations and identical source positions. 
Regarding excited-state contamination, we use the measurements of Table~\ref{tab:stat} at multiple $t_s$ values. 
This comes at no additional computational cost, as, by construction, disconnected contributions are evaluated at open sink time.

\vspace*{0.25cm}
\begin{table}[h!]
\begin{center}
\renewcommand{\arraystretch}{1.8}
\begin{tabular}{c|ccccc}
\hline
$P$ [GeV]  &  $\quad N_{\rm confs} \quad$ &  $\quad N_{\rm src} \quad $  &  $\quad N_{\rm dir}\quad  $ & $\quad N_{\rm meas}\quad $\\
\hline
0, 0.42, 0.83, 1.25, 1.67\,\,\, & 1,134 & 200 & 6 & 1,360,800\\
\hline
\end{tabular}
\vspace*{0.15cm}
\caption{\small{Total statistics for the calculation for each value of $P$. $N_{\rm confs}$ is the number of configurations, $N_{\rm src}$ the number of source positions, $N_{\rm dir}$ is the number of spatial directions for the Wilson line and $P$, and $N_{\rm meas}$ is the number of total measurements ($N_{\rm meas} = N_{\rm confs} \times N_{\rm src} \times N_{\rm dir}$). }}
\label{tab:stat}
\end{center}
\end{table}

The increased gauge noise is addressed by employing the stout smearing smoothing technique~\cite{Morningstar:2003gk} on the gauge links entering the gluon field strength tensor and the Wilson line. 
The stout smearing parameter is $\rho=0.129$~\cite{Alexandrou:2016ekb,Alexandrou:2020sml}, and the number of smearing steps is chosen independently in the gluon field strength tensor ($N^{\rm F}_{\rm stout}$) and the Wilson line ($N^{\rm W}_{\rm stout}$).
We apply a 4D smearing to the field strength tensor and a 3D smearing to the gauge links of the Wilson line. 
We have tested a 4D smearing in the Wilson line, obtaining compatible results after the double-ratio renormalization.
We calculate 25 combinations of $N^{\rm F}_{\rm stout}$ and $N^{\rm W}_{\rm stout}$, by using the values 0, 5, 10, 15, and 20 for each. 

Another technique to decrease the noise-to-signal ratio is to improve the overlap with the proton ground state. 
We apply momentum smearing~\cite{Bali:2016lva} for the three highest momentum boosts, $P=0.83,\, 1.25,\, 1.67$ GeV, which has been proven essential in suppressing the gauge noise in matrix elements with boosted hadrons and non-local operators~\cite{Alexandrou:2016jqi}.
We found the optimized value at $\xi=0.6$ for the momentum smearing parameter. 

\medskip
The evaluation of the quark matrix elements is an extension of the previous work of Ref.~\cite{Alexandrou:2021oih}, which obtained the quark PDFs within the quasi-distributions method with momenta $P=0.42,\, 0.83,\, 1.25$ GeV.
Here, we added $P=0,\, 1.67$ GeV, so that we obtain the reduced ITDs for the quarks, $\mathfrak{M}_q$, needed in Eq.~\eqref{eq:mix}. 
As for the gluon case, we implement the momentum smearing method and five stout smearing steps on the gluon fields of the Wilson line entering the operator. 
To reduce the stochastic noise coming from the low modes~\cite{Abdel-Rehim:2016pjw} in the calculation of the quark loops, we compute the first $N_{ev}=200$ eigenpairs of the squared twisted-mass Dirac operator. 
Then, the low-mode contribution to the all-to-all propagator can be exactly reconstructed, and the high-modes can then be evaluated with stochastic techniques, such as hierarchical probing~\cite{Stathopoulos:2013aci}. 
The latter allows for the reduction of the contamination of the off-diagonal terms in the evaluation of the trace of Eq.~\eqref{eq:loop}, up to a distance $2^k$, using Hadamard vectors as basis vectors for the partitioning of the lattice. 
Here, we use $k=3$ in four dimensions leading to 512 Hadamard vectors. 
In addition to the hierarchical probing, we make use of the \emph{one-end trick}~\cite{Abdel-Rehim:2013wlz,Alexandrou:2013wca} and fully dilute spin and color sub-spaces. More information can be found in Ref.~\cite{Alexandrou:2021oih}, as well as Refs.~\cite{Alexandrou:2020sml,Alexandrou:2019olr,Alexandrou:2019brg,Alexandrou:2018sjm}.

\begin{figure}[h!]
    \centering
    \includegraphics[scale=0.5]{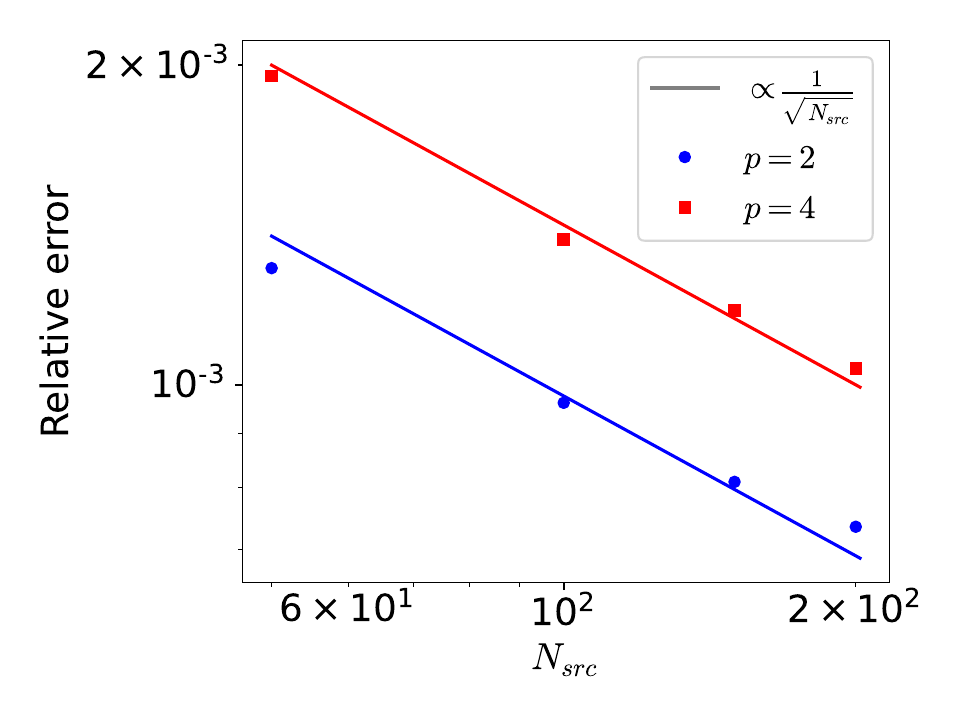}
    \vspace*{-0.5cm}
    \caption{\small{The relative error of the proton energy at momentum boost $P=\frac{2\pi}{L}p$ as a function of the source positions analyzed. As examples, we show $p=2$ (blue circles) and $p=4$ (red squares). The lines correspond to the $1/\sqrt{N_{src}}$ scaling.}}
    \label{fig:E_src}
\end{figure}

\medskip
On a large enough lattice, and given that the source positions are selected randomly, the autocorrelations become very small, and the data on multiple source positions on the same configuration can be considered statistically independent. 
To check for autocorrelations, we analyze different subsets of data for the two-point functions and extract the relative error on the energy, as shown in Fig.~\ref{fig:E_src} for two representative values of the momentum boost, $P=0.83$ GeV ($p=2$) and $P=1.67$ GeV ($p=4$).
We find that the statistical error of various quantities scales with the inverse square root of the number of source positions indicating uncorrelated data.  

\section{Results}
\label{sec:results}

\subsection{Gluon matrix elements and reduced ITDs}
\label{sec:MEs_and_DRs}

\begin{figure}[h!]
    \centering
    \includegraphics[scale=0.58]{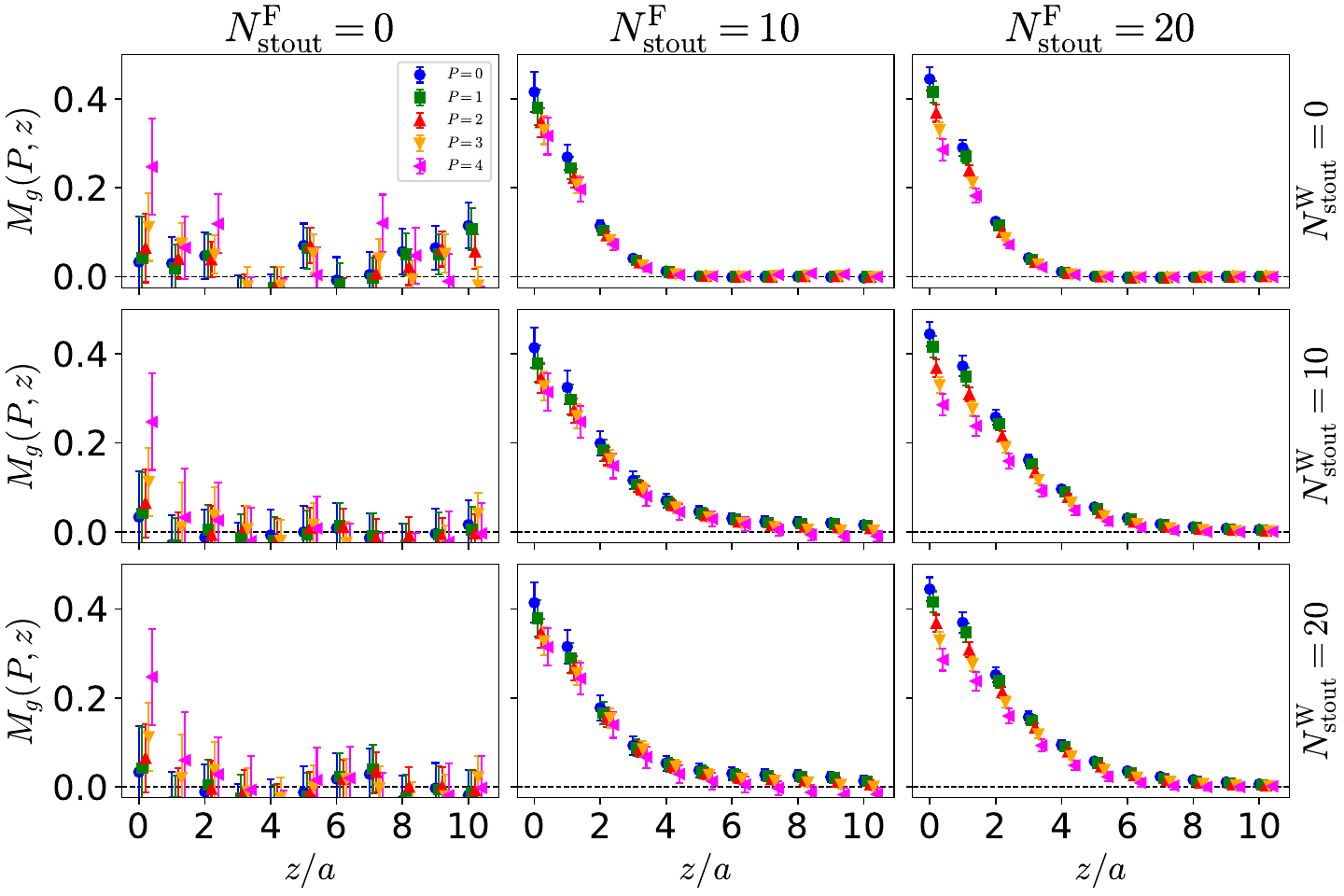}
    \vspace*{-0.5cm}
    \caption{\small{Stout smearing dependence of the bare matrix elements using $t_s=9a$. The left, center, and right columns correspond to $N^{\rm F}_{\rm stout}=0,\,10,\,20$, respectively. The top, center, and bottom rows correspond to $N^{\rm W}_{\rm stout}=0,\,10,\,20$, respectively. The data at momentum boost $P=\frac{2\pi}{L} p$ with $p=0,\,1,\,2,3,4$ are shown with blue circles, green squares, red up triangles, yellow down triangles, and magenta left triangles, respectively. }}
    \label{fig:M_NF}
\end{figure}

Before presenting the final bare matrix elements, it is useful to examine the effect of the stout smearing in terms of the signal quality. 
The stout smearing is extensively used in the calculation of non-local operators of Refs.~\cite{Alexandrou:2016jqi,Alexandrou:2018pbm,Alexandrou:2018eet, Alexandrou:2019lfo,Alexandrou:2020zbe,Alexandrou:2020qtt,Alexandrou:2020uyt,Alexandrou:2021bbo,Alexandrou:2020uyt,Alexandrou:2021oih} demonstrating the noise reduction. 
Also, in Ref.~\cite{Alexandrou:2019lfo}, we demonstrated the independence of the renormalized matrix elements from the level of smearing. 
However, the above statements regard quark bilinear operators, so similar tests are imperative for gluonic operators. 
As mentioned in the previous section, we construct the gluon matrix elements for 25 combinations of stout steps in the gluon field strength tensor and the Wilson line, that is $\{N^{\rm F}_{\rm stout},\,N^{\rm W}_{\rm stout}\} \in [0,\,20]$ in steps of 5. 
The bare matrix elements are shown in Fig.~\ref{fig:M_NF} for a subset of these combinations, which includes $\{N^{\rm F}_{\rm stout},\,N^{\rm W}_{\rm stout}\}=\{0,\,10,\,20\}$. 
All presented matrix elements have been evaluated at $t_s=9a$, which, as we will demonstrate below, is the one used in the final analysis. 
It is interesting to observe that the smearing on the field strength tensor has a bigger impact on the signal compared to the smearing on the Wilson line. 
For instance, the signal already improves significantly with $N^{\rm F}_{\rm stout}=10$ and $N^{\rm W}_{\rm stout}=0$. 

\begin{figure}[h!]
\vspace*{0.5cm}
    \centering
    \includegraphics[scale=0.58]{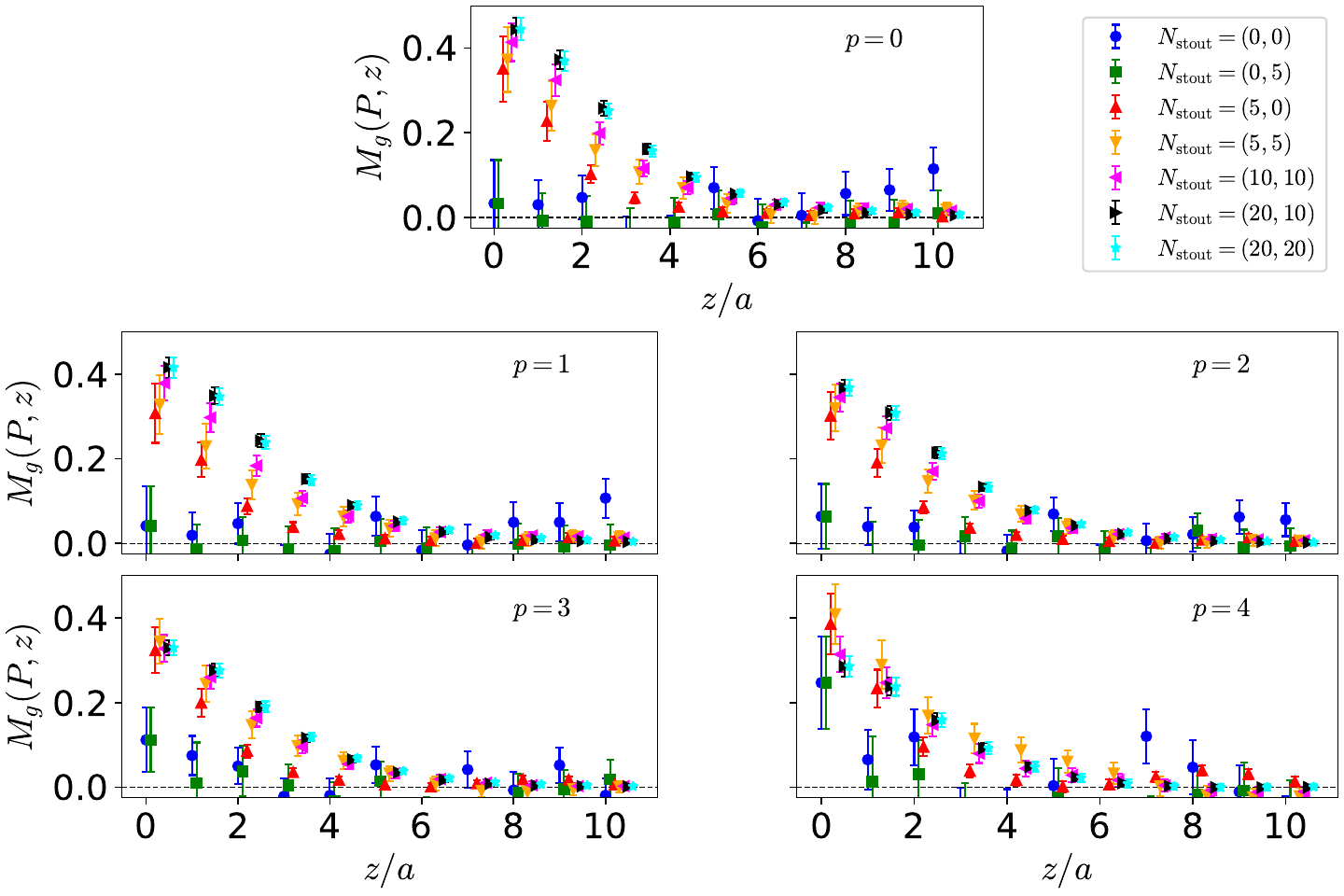}
    \vspace*{-0.5cm}
    \caption{\small{Stout smearing dependence of the bare matrix elements $M_g$. Results are given in coordinate space for the indicated combination $ N_{\rm stout} = (N^{\rm F}_{\rm stout},\,N^{\rm W}_{\rm stout}$). The data at momentum boost $P=\frac{2\pi}{L} p$ with $p=0,\,1,\,2,3,4$ are shown in the top, middle left, middle right, bottom left, and bottom right panels.}}
    \label{fig:M_NF_P}
\end{figure}
\begin{figure}[h!]
\centering
\includegraphics[scale=0.55]{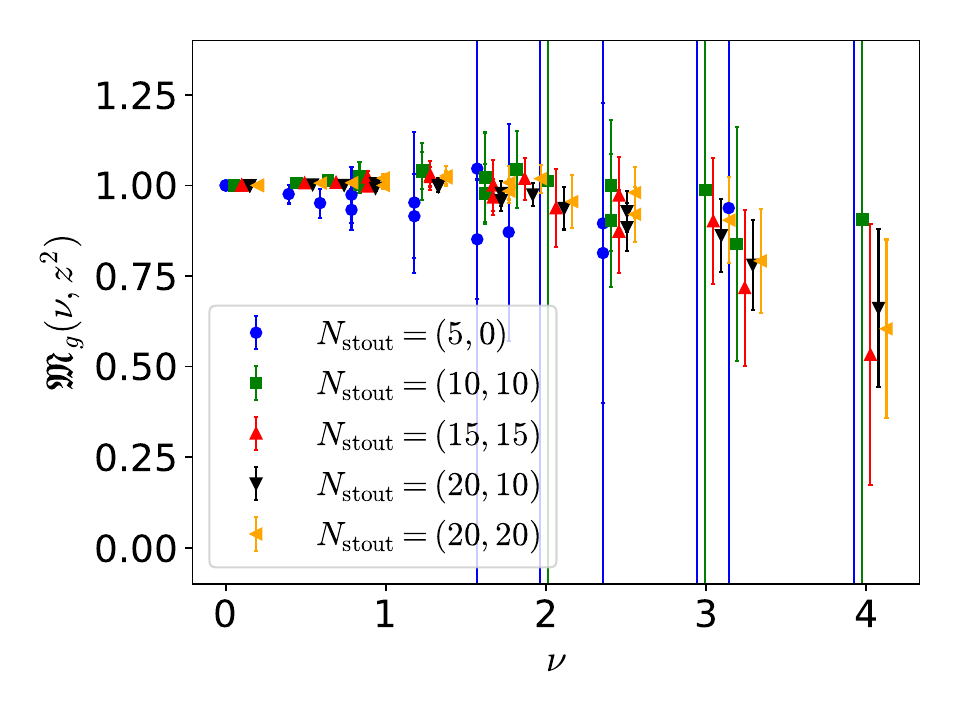}
\hspace*{-0.5cm}
\includegraphics[scale=0.55]{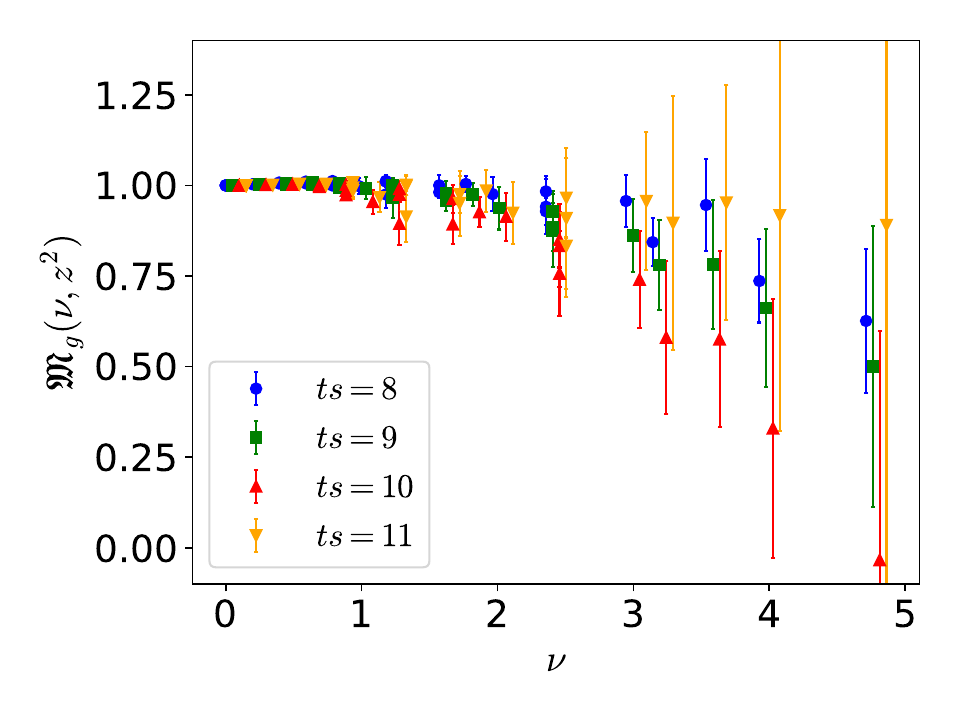}
\vspace*{-0.6cm}
\caption{\small{Left: Stout smearing number of steps dependence of the reduced ITDs at $t_s=9a$. Results for $(N^{\rm F}_{\rm stout},\,N^{\rm W}_{\rm stout})=(10,10),\,(15,\,15),\,(20,10),\,(20,20)$ are shown with blue circles, green squares, red up triangles, black down triangles, and orange left-pointing triangles, respectively. Right: Excited states in the reduced ITDs at $N^F_{\rm stout}=20$ and $N^W_{\rm stout}=10$. $t_s=8a,\,9a,\,10a,\,11a$ are shown with blue circles, green squares, red up triangles, and orange down triangles, respectively.}}
\label{fig:DR_stout_test}
\end{figure}

Comparing the effect of the stout smearing directly at each momentum can offer another qualitative understanding of signal improvement. 
In Fig.~\ref{fig:M_NF_P} we show selected cases of the $N^{\rm W}_{\rm stout}$ and $N^{\rm F}_{\rm stout}$ combinations, presented as $N_{\rm stout} = (N^{\rm F}_{\rm stout},\,N^{\rm W}_{\rm stout})$. 
As previously discussed, the stout smearing applied on the gluon fields of the field strength tensor is crucial to get a signal. 
In all values of $P$, further signal improvement is found as $N^{\rm F}_{\rm stout}$ and $N^{\rm W}_{\rm stout}$ increase. 
We observe a saturation at $(N^{\rm F}_{\rm stout},\,N^{\rm W}_{\rm stout})=(20,10)$, which we will use for the remainder of this analysis. 
In Fig.~\ref{fig:DR_stout_test}, we will examine the effect of the mixing in the pseudo ITDs.

\begin{figure}[h!]
  \includegraphics[scale=0.58]{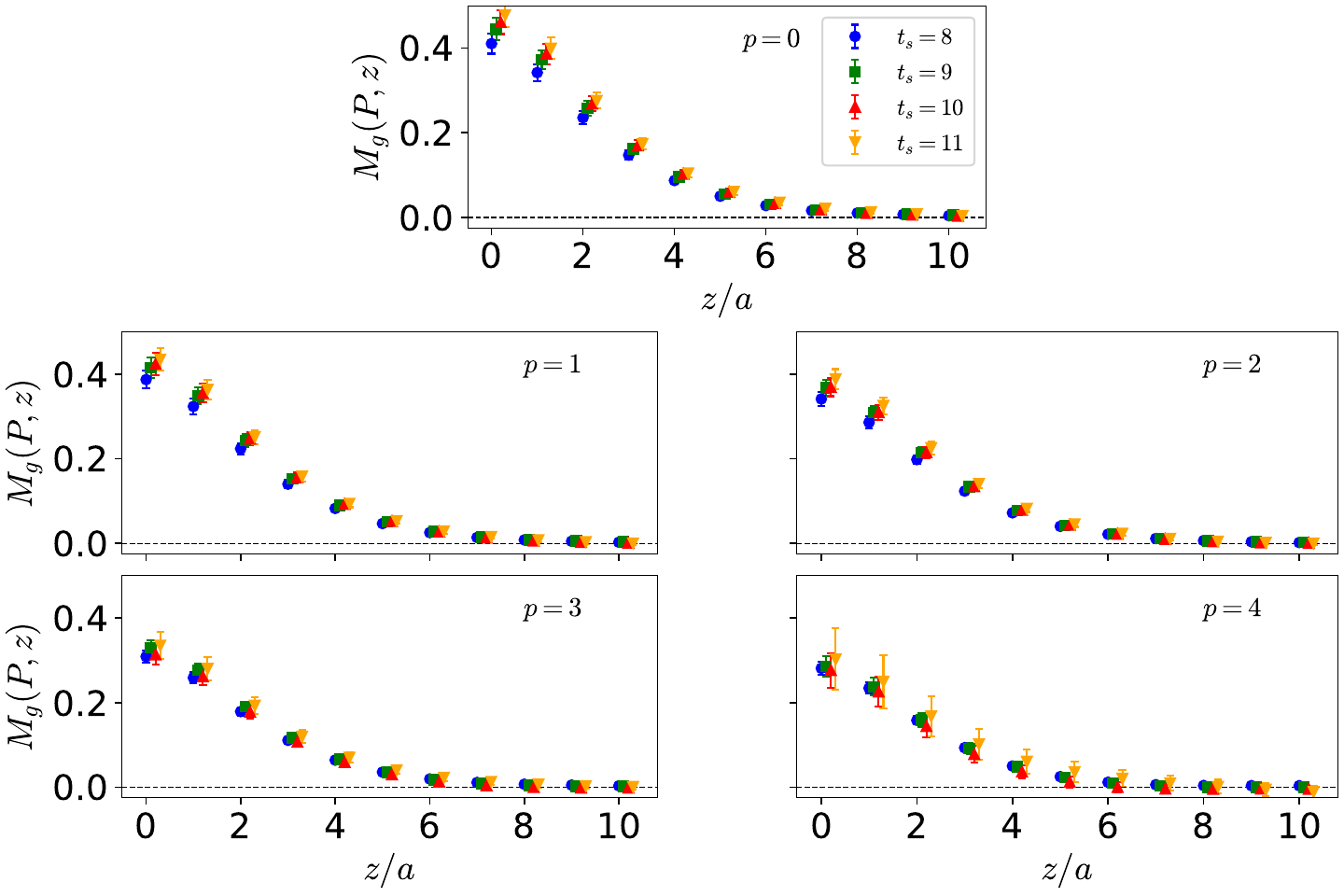}
    \caption{\small{Source-sink time separation dependence of the bare matrix elements at each momentum boost. We use $N^{\rm W}_{\rm stout}=10$ and $N^{\rm F}_{\rm stout}=20$ in all cases. The results for $t_s=8a,\,9a,\,10a,\,11a$ are shown in blue circles, green squares, red up triangles, and orange down triangles, respectively. The data at momentum boost $P=\frac{2\pi}{L} p$ with $p=0,\,1,\,2,3,4$ are shown in the top, middle left, middle right, bottom left, and bottom right panels, respectively. }}
    \label{fig:ME_ts_comp}
\end{figure}

\newpage
In this work, we also examine excited-states effects using the preferred setup for the stout smearing, $N^{\rm F}_{\rm stout}=20$, $N^{\rm W}_{\rm stout}=10$. 
Fig.~\ref{fig:ME_ts_comp} shows the matrix elements at four values of the source-sink time separation, that is, $t_{s}=8a,\, 9a,\,10a,\,11a$. 
For $P=$ 0, 0.42, and 0.83 GeV, there is an indication of excited-states effects at $t_s=8a$, which differs from $t_s=10a$ and $t_s=11a$. 
The effect is visible mainly due to the high statistical accuracy of the data. 
For higher momenta, all matrix elements are compatible within uncertainties, which are enhanced compared to the lower momenta. 
Therefore, $t_s=9a$ is favorable, as it is consistent with $t_s=10a$ and $t_s=11a$ in all cases, while good signal is maintained. 
Below, we will also consider excited-states effects in the reduced ITDs (see, e.g., Fig.~\ref{fig:DR_stout_test}).

\begin{figure}[h!]
    \centering
    \includegraphics[scale=0.5]{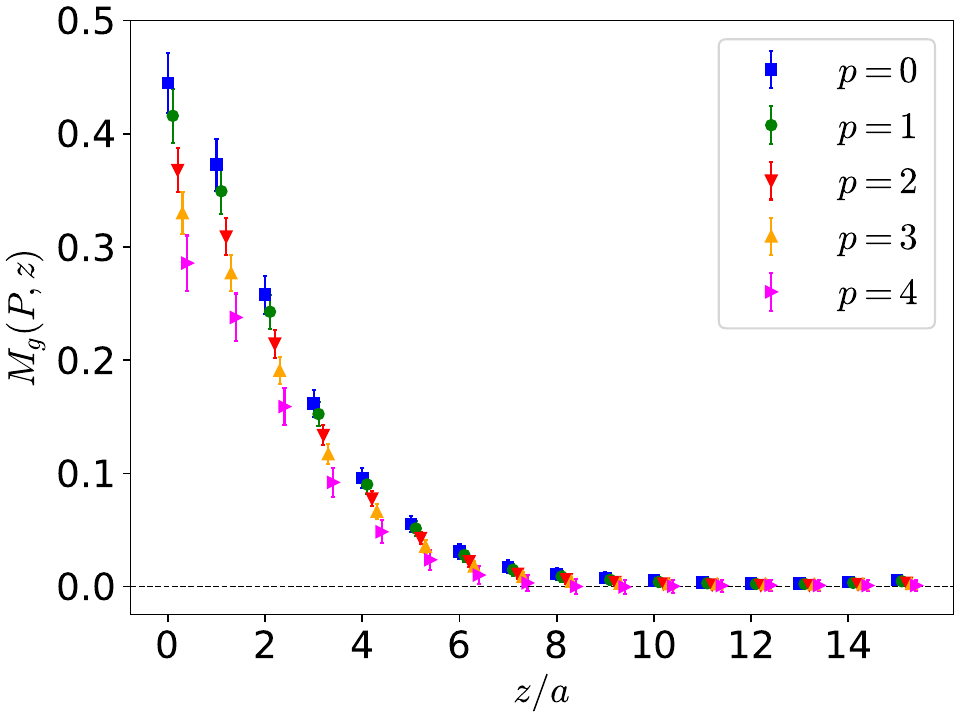}
    \vspace*{-0.5cm}
    \caption{\small{Matrix elements of Eq.~\eqref{eq:gluon_oper} as a function of the length of the Wilson line, $z/a$. The data at momentum boost $P=\frac{2\pi}{L} p$ with $p=0,\,1,\,2,3,4$ are shown with blue squares, red circles, green downward-pointing triangles, yellow upward-pointing triangles, and magenta rightward-pointing triangles, respectively. }}
    \label{fig:gluon_matrix}
\end{figure}
To summarize the presentation of the bare matrix elements, we compare in Fig.~\ref{fig:gluon_matrix} the data for all values of the momentum boost using $t_s=9 a$,  $N^{\rm F}_{\rm stout}=20$, and $N^{\rm W}_{\rm stout}=10$. 
The $P$ dependence of the data is as observed in the quark case, that is, the signal quality decreases. 
We find that the relative error at $z=0$ for $P=0$ is about 6\%, while for $P=1.67$ GeV, the error becomes close to 9\% despite the same statistics. 
In all cases, we find that the matrix elements decay to zero at about $z=8a$.

The matrix elements of Fig.~\ref{fig:gluon_matrix} are the core of our calculation and are used to construct the double ratio of Eq.~\eqref{eq:double_ratio}. 
We note that systematic uncertainties might affect the matrix elements and pseudo ITDs differently due to possible correlations between the ratios in the numerator and/or denominator. 
Thus, investigating systematic effects, such as excited states and stout smearing, in the ratio of Eq.~\eqref{eq:double_ratio} is important. 
Since the pseudo ITD, i.e.\ the double ratio, serves as a renormalization prescription, it should be independent of the number of smearing steps. 
We examine the validity of this argument, and a summary is shown in Fig.~\ref{fig:DR_stout_test}. 
Due to the large uncertainties of certain combinations of smearing steps, their inclusion in the plot is not meaningful, as their errors cover the whole range of the plot. As can be seen, all combinations of $N^{\rm F}_{\rm stout}$ and $N^{\rm W}_{\rm stout}$ are in full agreement within errors, demonstrating that the pseudo ITDs can be extracted from any of these combinations.

\begin{figure}[h!]
    \centering
    \includegraphics[scale=0.55]{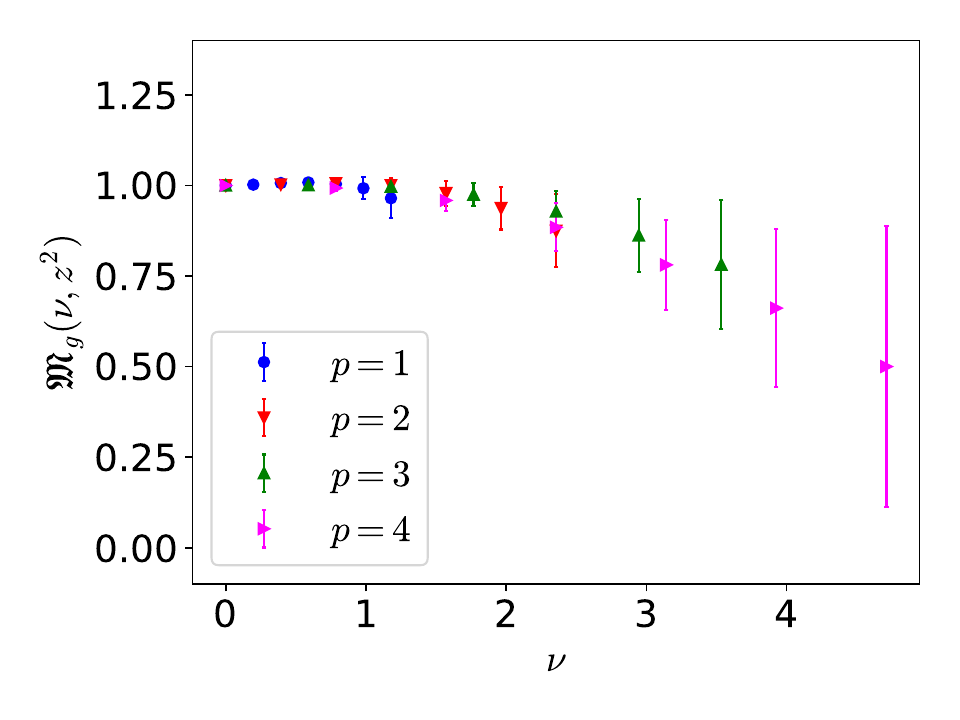}
    \vspace*{-0.6cm}
    \caption{\small{Final results for the reduced-ITD at $t_s=9a$, $N^F_{\rm stout}=20$ and $N^W_{\rm stout}=10$. We show data with all values of $P=\frac{2\pi}{L}p$ and $z$ up to $6a\sim0.56$ fm. Data for $p=1,\,2,\,3,\,4$ are shown with blue circles, red down triangles, green up triangles, and magenta right triangles, respectively.}
    \label{fig:DR_final}}
\end{figure}
We study excited-states effects in $\mathfrak{M}_g$, as shown in Fig.~\ref{fig:DR_stout_test} for four values of the source-sink time separation, that is $t_s=8a,\,9a,\,10a,\,11a$. 
The increase of the statistical error is sizeable between $t_s=9a$ and $t_s=11a$ and the signal is lost at $t_s=12a$; the latter is not shown here. 
Overall, we find that both $t_s=8a$ and $t_s=9a$ are good options for these data. 
Nevertheless, we choose $t_s=9a$ for a more conservative estimate. 
For completeness, we show in Fig.~\ref{fig:DR_final} the double ratio for all values of $P$ corresponding to $t_s=9a$, $N^F_{\rm stout}=20$ and $N^W_{\rm stout}=10$. 
We note that each value of $\nu$ is constructed from all possible combinations of available $z$ and $P$, but we constrain $z$ up to $6a\sim0.56$ fm. 
This leads to $\nu_{\rm max}\sim5$.
Comparing all combinations of $P$ and $z$ at a given $\nu$ allows one to comment on the effect of $P$ dependence. 
We find that dependence on $P$ is within the statistical errors for up to $z=6a$. 
Thus, $\mathfrak{M}_g$ can be described by a smooth function in terms of the Ioffe time, which allows for a controlled interpolation. 
The latter is needed for the scale evolution and matching procedure, as discussed below.

\begin{figure}[h!]
    \centering
    \includegraphics[scale=0.55]{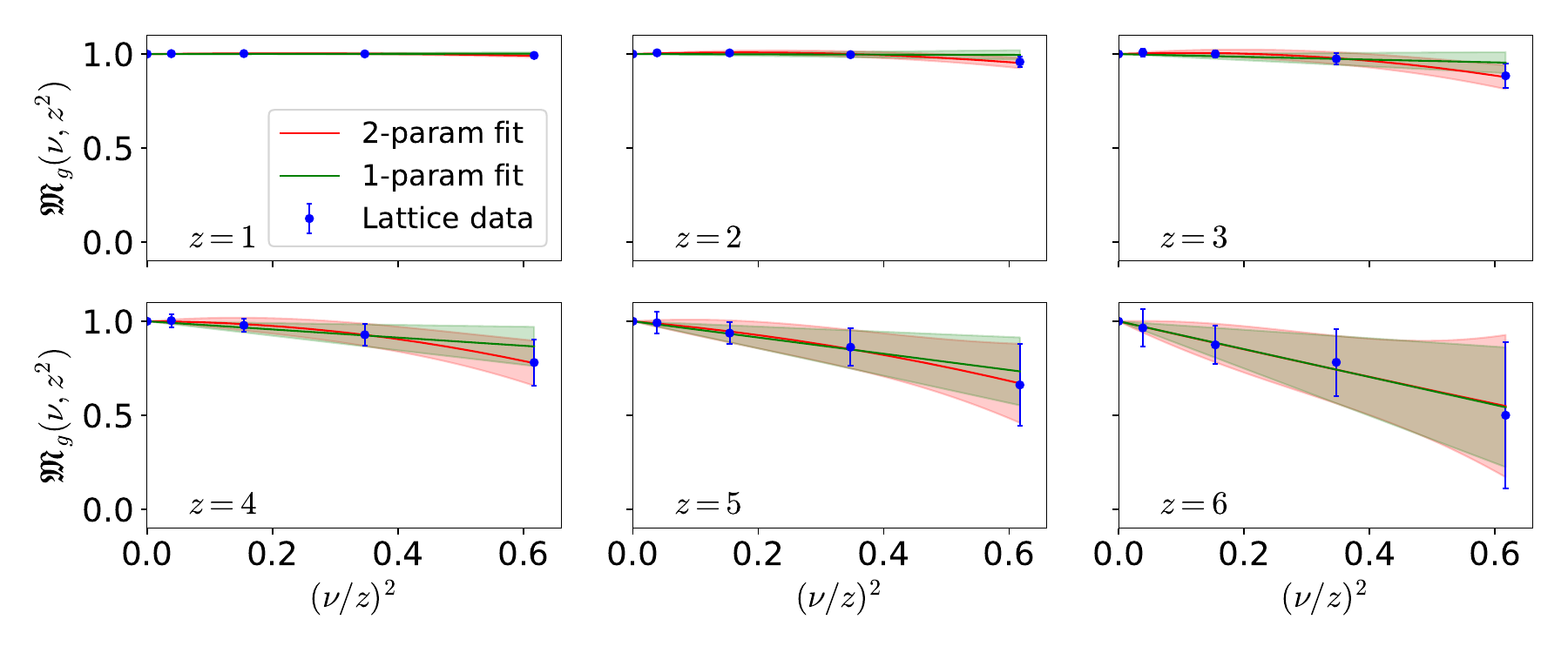}
    \vskip -0.5cm
    \caption{\small{Lattice data of the reduced ITDs for $z=1a - 6a$ (blue points) and their interpolation at fixed $z^2$ using a first order (green bands) and second order polynomial fit (red bands). }}
    \label{fig:DR_interpolation}
\end{figure}

The reduced ITDs are interpolated in terms of $\nu$ at each value of $z^2$, so that one obtains a continuous function in $\nu/z$, which is needed for the matching procedure. 
Having five values of $\nu$ at a fixed $z^2$, one can test different parametrizations of the $\nu$ dependence. 
Here, we test a linear and a second-order polynomial fit, which can be seen in Fig.~\ref{fig:DR_interpolation} for selected values of $z$. We find that the two fits are compatible and choose the polynomial fit to proceed.

\subsection{Reconstruction of the gluon PDF}
\label{sec:no_mixing}

\begin{figure}[h!]
    \centering
    \includegraphics[scale=0.44]{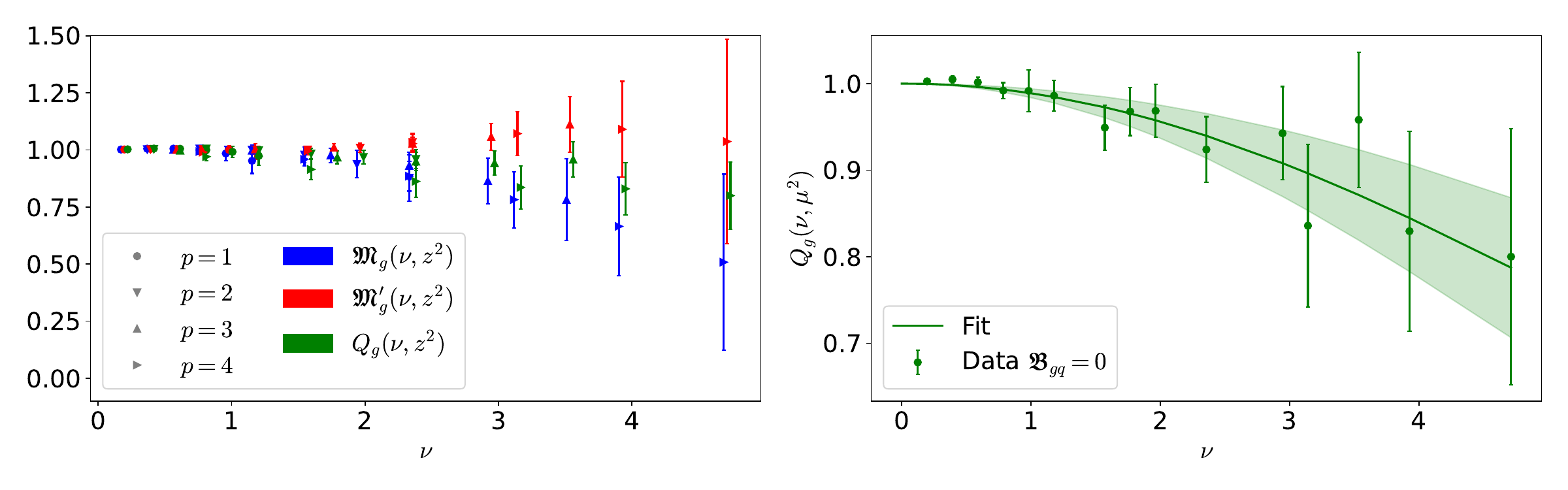}
    \vspace*{-0.85cm}
    \caption{\small{Left: The reduced (blue), evolved (red), and matched (green) ITDs (no mixing with the quark singlet PDF) shown for momentum boosts $p=1$ (circles), $p=2$ (down-pointing triangles), $p=3$ (up-pointing triangles), and $p=4$ (right-pointing triangles), where $p$ is defined through $P=\frac{2\pi}{L}p$. Right: The light-cone ITDs (no mixing) averaged over $(P,z)$ pairs giving the same Ioffe time and the fitting band of the fitting ansatz reconstruction. }}
    \label{fig:evolved_ITD}
\end{figure}

In general, the extraction of light-cone ITDs from reduced ITDs contains combining effects of three functions, the $\mathfrak{B}_{gg}$, $L$ and $\mathfrak{B}_{gq}$ kernels, as given in Eq.~\eqref{eq:itd_equation}. 
That is, one must apply the evolution to a renormalization scale of choice ($\mu$), convert the data to light-cone ITDs in the $\rm \overline{MS}$ scheme, and eliminate the mixing with the quark-singlet PDF. 
Here, we chose 2 GeV for the renormalization scale, as commonly used in global analyses. 
In all previous calculations of the gluon PDF, the mixing with the quark-singlet case has been ignored due to the lack of lattice results for the latter, as it requires information from disconnected contributions, which are computationally very expensive. 
Here, we extend the calculation of Ref.~\cite{Alexandrou:2021oih} to include all values of $P$ implemented in this work, which allows us to eliminate the mixing by considering $\mathfrak{B}_{gq}$.
To demonstrate the effect of the mixing, we first apply $\mathfrak{B}_{gg}$ and $L$, but ignore $\mathfrak{B}_{gq}$. 
The resulting evolved and matched ITDs are shown in Fig.~\ref{fig:evolved_ITD}. 
We find that the scale evolution increases the values of the evolved ITDs ($\mathfrak{M}'$) relative to those of the reduced ITDs ($\mathfrak{M}$), while the matching has the opposite effect than the evolution and brings the light-cone ITDs ($Q$) closer to the reduced ITDs, making them consistent with the latter within error bars. 
Such behavior is also observed in the case of quark PDFs (see, e.g., Refs.~\cite{Bhat:2020ktg,Bhat:2022zrw}). 
We note that the dependence on the individual $P$ and $z$ is minimal for all three functions, $\mathfrak{M}$, $\mathfrak{M}'$, and $Q$, as the values from different $(P,z)$ pairs fall on a universal curve. 
In the right panel of Fig.~\ref{fig:evolved_ITD}, we show the matched ITDs, $Q(z^2,\mu^2)$, where we average over the $(P,z)$ pairs for a given value of the Ioffe time. 

To extract the $x$-dependence of the gluon PDF, we use the fitting reconstruction and follow the procedure discussed in Sec.~\ref{sec:setup}. 
As can be seen in Eq.~\eqref{eq:ansatz}, one cannot isolate the gluon PDF, because it appears normalized with the gluon momentum fraction. 
The latter has not been extracted on the ensemble under study, so we use, instead, the lattice results of Ref.~\cite{Alexandrou:2020sml}. 
The aforementioned calculation used an ensemble that has the same gluon and fermion action as this work, but different lattice parameters. 
In particular, the lattice spacing is 0.08 fm, and the pion mass is 139 MeV. 
The reported value for the gluon momentum fraction is $\langle x \rangle^{\rm \overline{MS},\,2GeV}_g=0.427(92)$, which we use below.

Another input of the reconstruction procedure is the value of $\zmax$.
We tested $\zmax=5a,\,6a,\,7a$, which correspond to $\zmax=0.47,\,0.56,\,0.66$ fm, respectively.
In this subsection, we show results for $\zmax=6a$ and we demonstrate the independence of the results on $\zmax$ in the case where mixing with the quark singlet is eliminated (see next subsection).
The conclusions of this test fully pertain also to the case with the mixing neglected.

\begin{figure}[h!]
    \centering
    \includegraphics[scale=0.44]{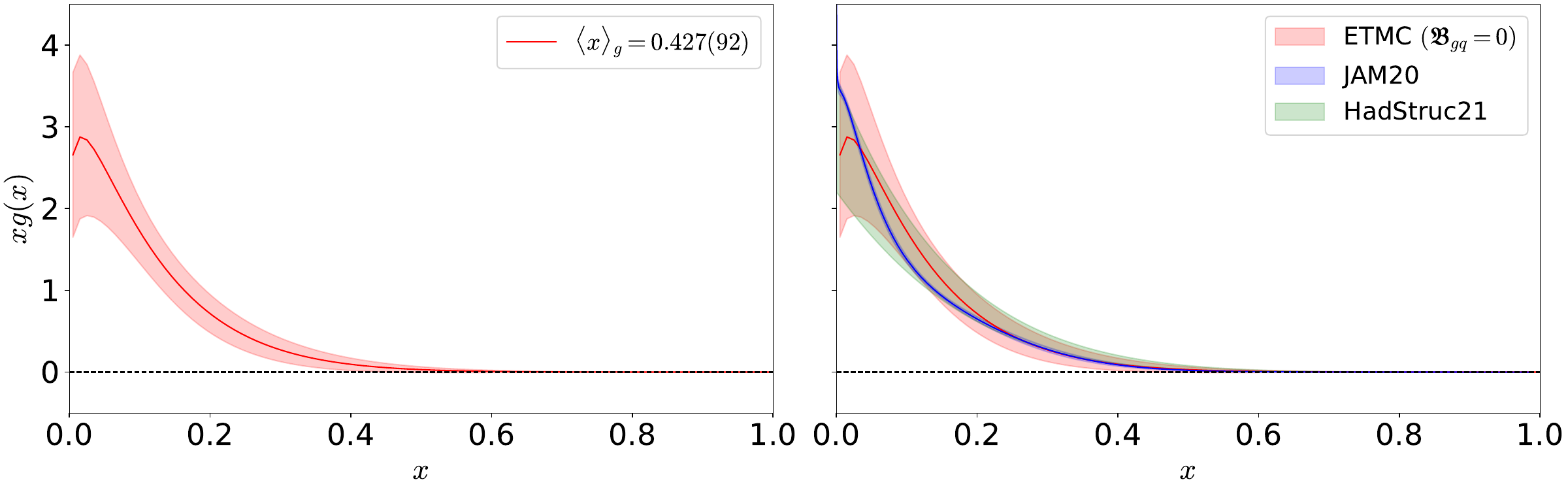}
        \vskip -0.35cm
    \caption{\small{Left: The reconstructed gluon PDF without mixing elimination. Right: A comparison of our results from the left panel (red), the lattice results of HadStruc~\cite{HadStruc:2021wmh} (green), and the global analysis of JAM20~\cite{Moffat:2021dji} (blue). Results are shown in the $\mathrm{\overline{MS}}$ scheme at a scale of 2 GeV.}}
    \label{fig:reconstructed_PDF}
\end{figure}

Using the above value of $\langle x \rangle^{\rm \overline{MS},\,2GeV}_g$ and $\zmax=6a$, we obtain the gluon PDF, which is given in the left panel of Fig.~\ref{fig:reconstructed_PDF}. 
The corresponding fitted ITDs are shown in the right panel of Fig.~\ref{fig:evolved_ITD}.
We remind the reader that we have not yet considered the mixing with the quark-singlet PDF; this will be addressed in the next subsection. 
In the right panel of Fig.~\ref{fig:reconstructed_PDF}, we compare our final results to the lattice results of HadStruc~\cite{HadStruc:2021wmh}, in which the gluon - quark singlet mixing has not been considered. 
HadStruc used an ensemble of $N_f=2+1$ clover Wilson fermions with stout-link smearing and the Symanzik-improved gauge action. The ensemble has the same volume and lattice spacing as this work. However, their pion is heavier, namely $m_\pi=358$ MeV.
Their source-sink time separation is also $9a$, which is the same as the value used here. 
In general, our results are consistent with the ones from HadStruc. 
It is worth noting that the reconstruction performed by HadStruc includes values of Ioffe time up to $\nu_{\rm max}=7.07$, while our reconstruction includes up to a maximum Ioffe time of $\nu_{\rm max}=4.71$ ($\zmax=6a$). 
The smaller statistical error of HadStruc may possibly be attributed to two factors: (a) the use of the distillation method~\cite{HadronSpectrum:2009krc}; (b) the higher pion mass compared to this work. 
In Fig.~\ref{fig:reconstructed_PDF}, we also compare the lattice data to the global analysis of JAM20~\cite{Moffat:2021dji}.
As can be seen, all results are in full agreement within errors. 
We note that all comparisons are qualitative, as the lattice results are obtained on a single ensemble with different lattice formulations. 
Nevertheless, the agreement between lattice results and global analysis is very promising.

\subsection{Elimination of mixing with quark-singlet PDF}
\label{sec:mixing}

In this section, we provide, for the first time, the quark-singlet PDF using the pseudo-distribution method.
This is a continuation of the work of Ref.~\cite{Alexandrou:2021oih}, which used a subset of the data of Table~\ref{tab:stat} to obtain the quark PDFs within the quasi-distributions method.
The quark-singlet PDF will be used to eliminate the mixing with the gluon contribution using the matching formalism of Ref.~\cite{Ji:2022thb}. 
In principle, with our data for the quark and gluon PDFs, we can also obtain the quark-singlet PDFs without mixing. However, Refs.~\cite{Balitsky:2019krf,Balitsky:2021bds} only provide the components of the mixing kernel that are relevant to the gluon PDF, that is, $\mathfrak{B}_{gg}$ and $\mathfrak{B}_{gq}$. 
While the complete $2\times2$ matching kernel is presented in Ref.~\cite{Ji:2022thb}, it corresponds to a different definition of the gluon operator than the one we use in this work, so we are not able to apply it here.

\begin{figure}[h!]
    \centering
    \includegraphics[scale=0.44]{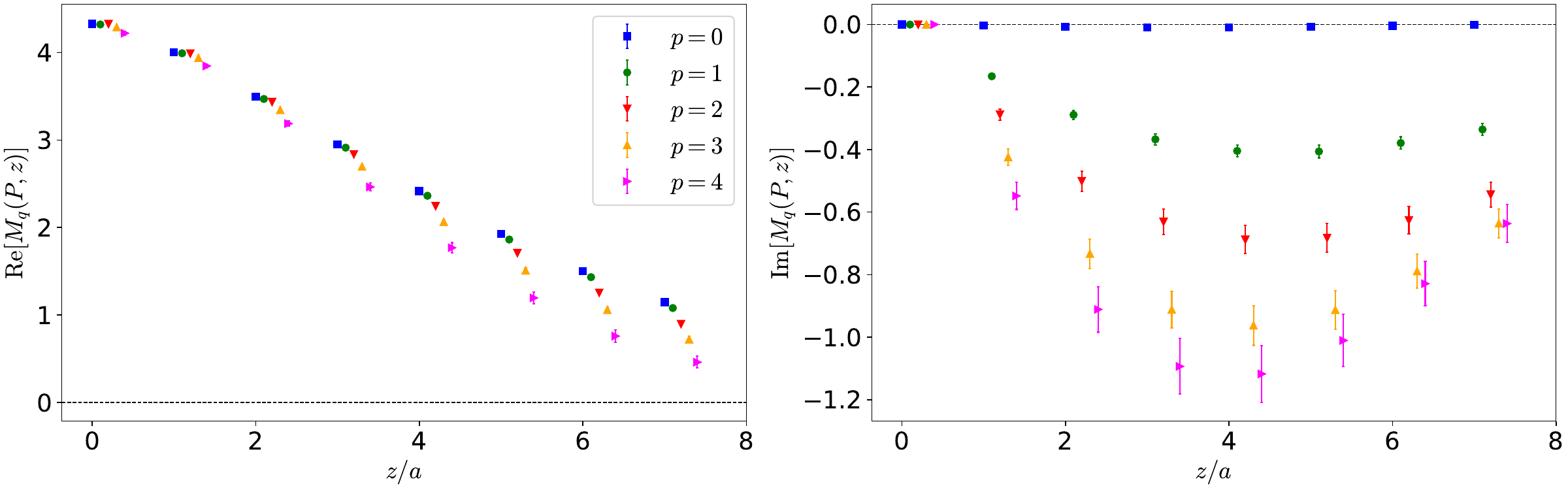}
    \vspace*{-0.5cm}
    \caption{\small{Bare matrix elements for the quark-singlet case as a function of the length of the Wilson line, $z/a$. The data at momentum boost $P=\frac{2\pi}{L} p$ with $p=0,\,1,\,2,3,4$ are shown with blue squares, green circles, red downward-pointing triangles, yellow upward-pointing triangles, and magenta rightward-pointing triangles, respectively. }}
    \label{fig:quark_matrix}
\end{figure}

\begin{figure}[h!]
    \centering
    \includegraphics[scale=0.44]{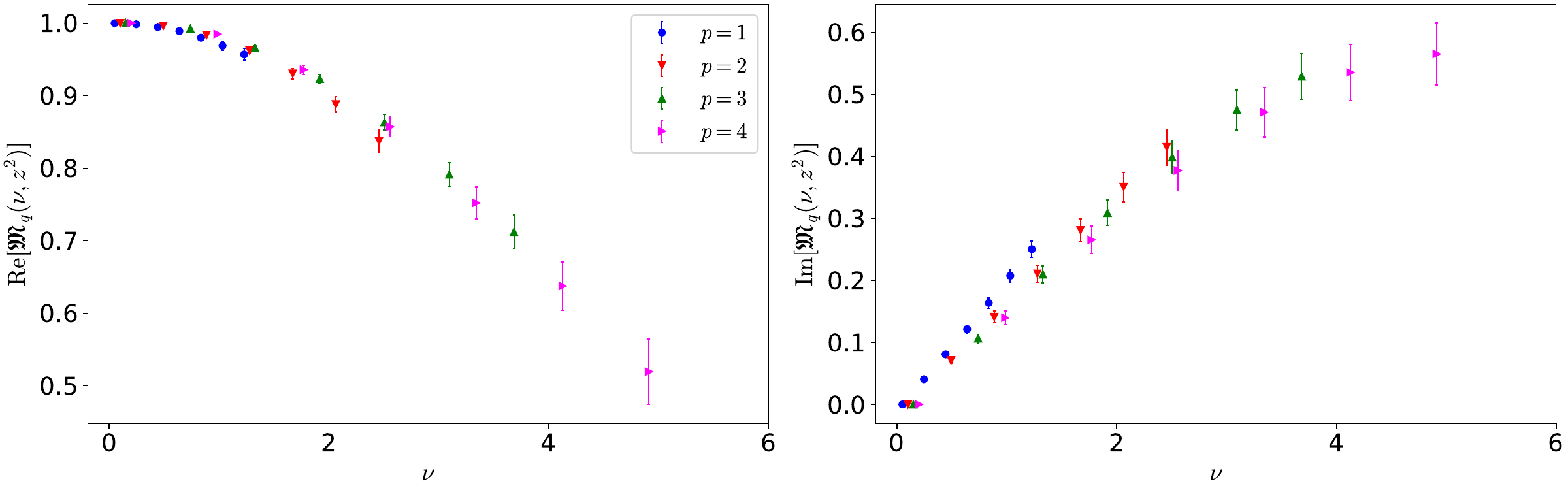}
    \caption{\small{Quark-singlet reduced-ITD for $z$ up to $6a\sim0.56$ fm. Data at momentum boost $P=\frac{2\pi}{L} p$ with $p=1,\,2,\,3,\,4$ are shown with blue circles, red down triangles, green up triangles, and magenta right triangles, respectively.}}
    \label{fig:DR_final_quark}
\end{figure}

First, let us present the bare quark matrix elements for the singlet combination $u+d+s$.
The matrix elements contain all kinematic factors, so they can be compared directly at $z=0$ for different values of $P$.
As seen in Fig.~\ref{fig:quark_matrix}, the data are consistent at $z=0$. 
This is expected theoretically, because $z=0$ is directly related to $\langle x \rangle$, which is independent of the kinematic frame. 
As $z$ increases, we find that the behavior with the increase of $P$ is as expected. 
That is, the real part of the matrix element falls faster, while the imaginary part is enhanced.
The corresponding quark reduced ITDs are shown in Fig.~\ref{fig:DR_final_quark}. In the real part, we find an agreement between the different $P$ and $z$ combinations corresponding to the same value of $\nu$. 
Some difference is observed in the imaginary part for $\{p,\,z/a\}=\{1,\,4\}$ as compared to $\{p,\,z/a\}=\{2,\,2\}$ and $\{p,\,z/a\}=\{4,\,1\}$ (where $P=\frac{2\pi}{L}p$). 
Similarly, $\{p,\,z/a\}=\{1,\,6\}$ deviates from $\{p,\,z/a\}=\{2,\,3\}$, and $\{p,\,z/a\}=\{3,\,2\}$. 
However, the momenta with $p>1$ are in agreement within errors. 

\begin{figure}[h!]
    \centering
    \includegraphics[scale=0.44]{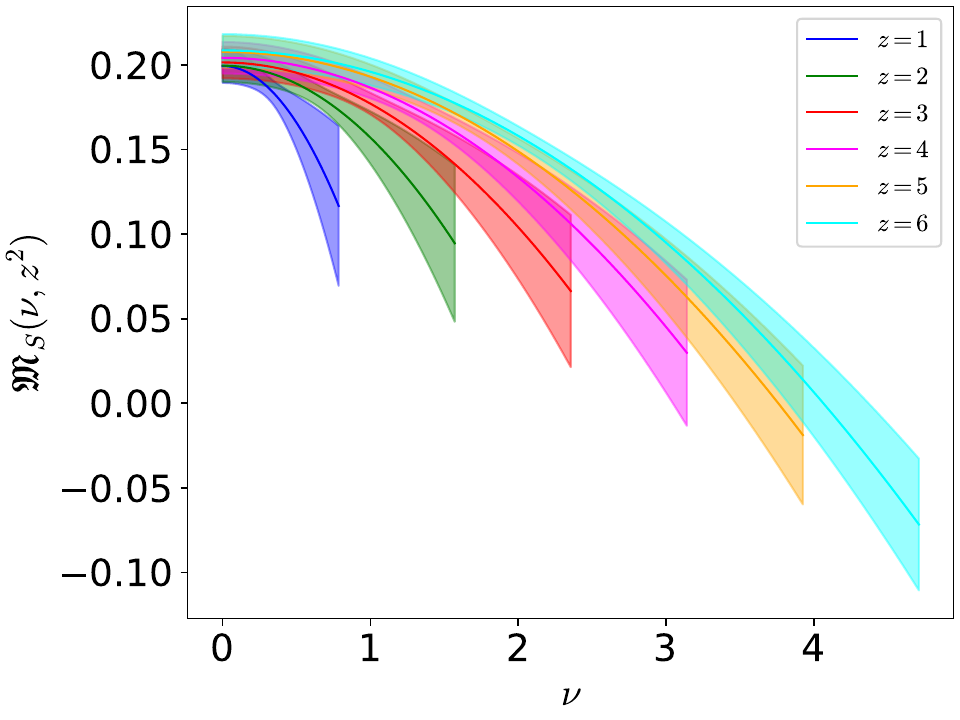}
    \vspace*{-0.25cm}    
    \caption{\small{The $\nu$-derivative of the imaginary part of $\mathfrak{M}_q$, $\mathfrak{M}_S$ as a function of $\nu$, as obtained from $z=1,\,2,\,3,\,4,\,5,\,6$ shown in blue, green, red, magenta, orange, and cyan bands.}}
    \label{fig:Ms}
\end{figure}

\begin{figure}[h!]
    \centering
    \includegraphics[scale=0.45]{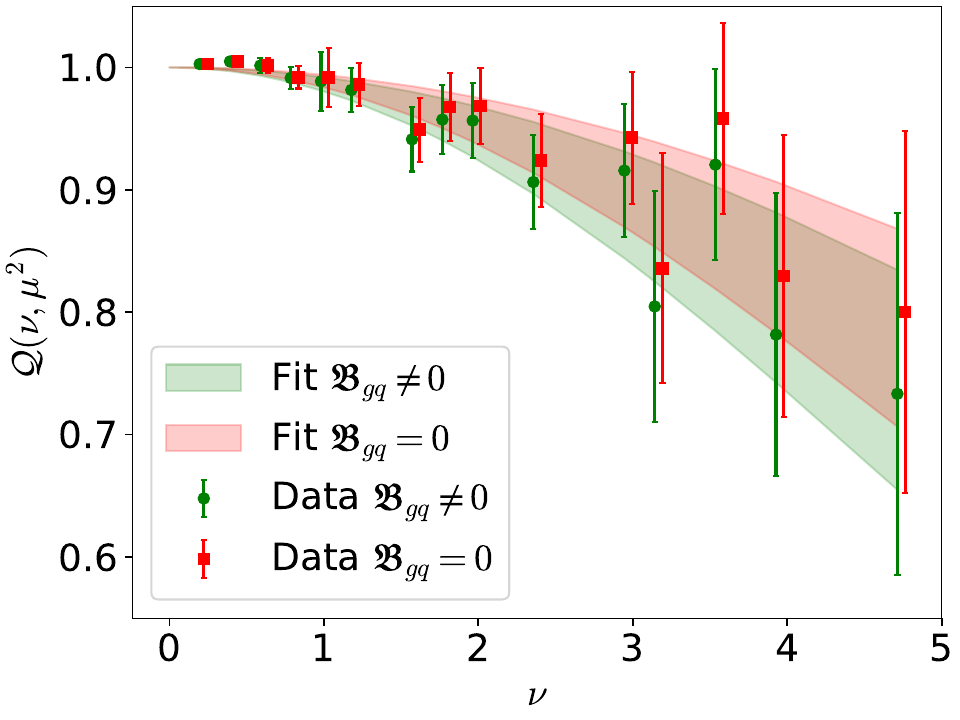}
    \vspace*{-0.25cm}
    \caption{\small{Comparison of the light-cone ITD before ($\mathfrak{B}_{gq}=0$, shown in red) and after ($\mathfrak{B}_{gq}\ne0$, shown in green) the elimination of the mixing with the quark-singlet case. The bands correspond to the fits of the lattice data. }}
    \label{fig:ITD_no_mixing}
\end{figure}

We use the above quark-singlet reduced ITDs to eliminate the mixing in the light-cone gluon ITDs. 
In particular, only $\mathfrak{M}_S$ --the $\nu$-derivative of the imaginary part of $\mathfrak{M}_q$-- enters the matching formalism, as explained in Sec.~\ref{sec:setup}.
For completeness we show $\mathfrak{M}_S$ in Fig.~\ref{fig:Ms}.
Finally, the resulting effect of the mixing is shown in Fig.~\ref{fig:ITD_no_mixing} by comparing the gluon ITDs before ($\mathfrak{B}_{gq}=0$) and after ($\mathfrak{B}_{gq}\ne0$) the elimination of the mixing with the quark-singlet. 
The fitting bands from the $x$-dependence reconstruction procedure are also shown.
The main finding is that the gluon ITDs move slightly towards lower values, with the mixing being well within the occurring statistical uncertainties. 

\begin{figure}[h!]
    \centering
    \includegraphics[scale=0.44]{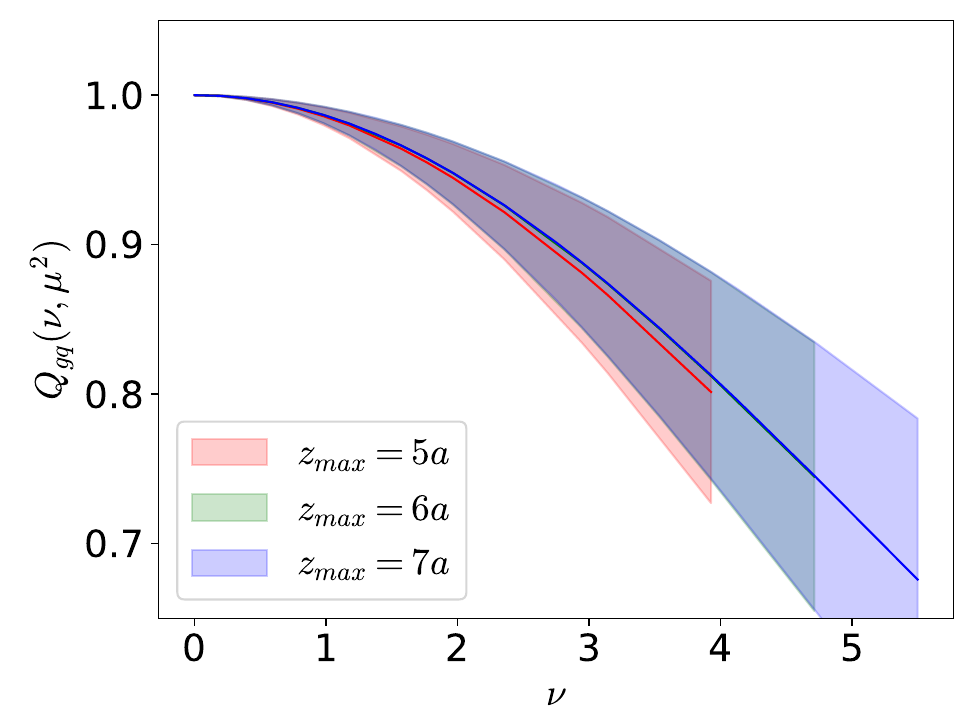}
\vspace*{-0.35cm}    
    \caption{\small{Comparison of fitted ITDs at different values of the maximum $z$ entering the reconstruction fit. $\zmax=5a,\,6a,\,7a$ are indicated by red, green, blue bands, respectively.}}
    \label{fig:ITD_z_max}
\end{figure}

As hinted in the previous subsection, we also establish the robustness of the results against the choice of the value of $\zmax$, using $\zmax=5a,\,6a,\,7a$, see Fig.~\ref{fig:ITD_z_max}.
We find a small difference between $\zmax=5a$ and $\zmax=6a$, but the effect is significantly smaller than the statistical uncertainties. 
The difference between $\zmax=6a$ and $\zmax=7a$ is almost negligible, and the two bands cannot be visually distinguished in the figure.
That is, the addition of $z=7a$ points does not influence the reconstruction, mainly due to their large statistical errors. 
Thus, our choice of $\zmax=6a$ is validated at this level of data precision and given the compatibility of results for different $\zmax$, it is not necessary to assign a reconstruction-related systematic uncertainty to our results.

\begin{figure}[h!]
    \centering
    \includegraphics[scale=0.42]{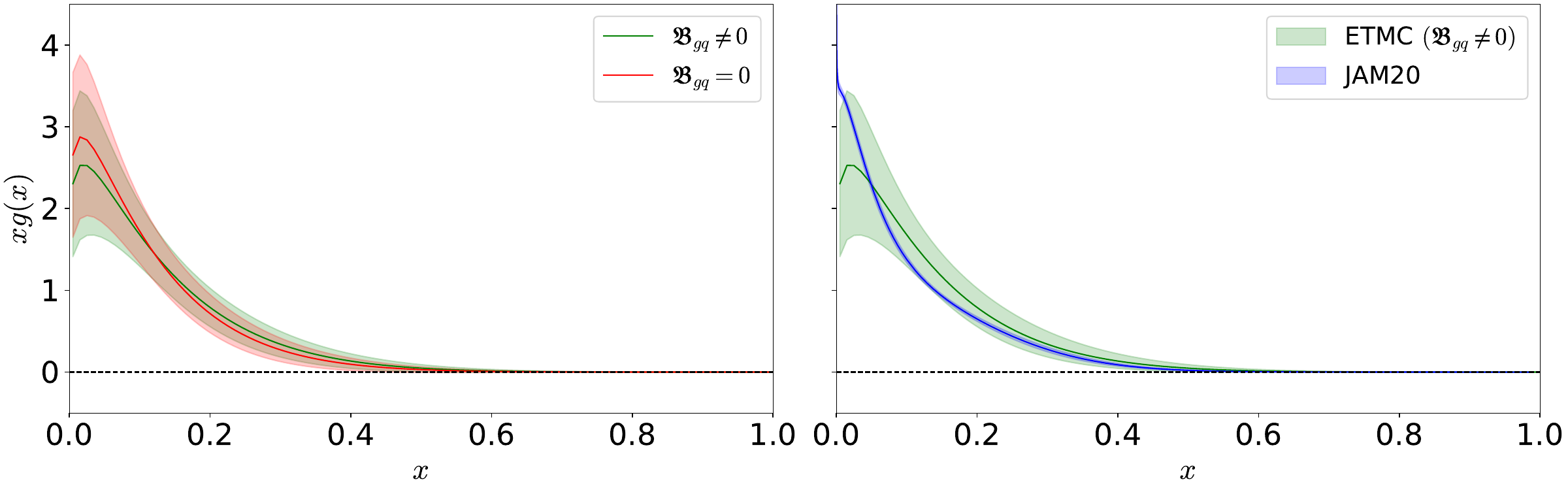}
        \vspace*{-0.25cm}
    \caption{\small{Left: The reconstructed gluon PDF before ($\mathfrak{B}_{gq}=0$) and after ($\mathfrak{B}_{gq}\ne0$) the elimination of the mixing with the quark-singlet PDF. Right: A comparison of our final results ($\mathfrak{B}_{gq}\ne0$) and the global analysis of JAM20~\cite{Moffat:2021dji}. Results are shown in the $\mathrm{\overline{MS}}$ scheme at a scale of 2 GeV.}}
    \label{fig:final}
\end{figure}

For completeness, we also present the effect of the mixing in the $x$-dependent gluon PDF, as seen in the right panel of Fig.~\ref{fig:final}. 
The conclusion is consistent with Fig.~\ref{fig:ITD_no_mixing}, as the effect of the mixing is smaller than the statistical uncertainties. 
In the left panel of Fig.~\ref{fig:final}, we show our final results together with JAM20~\cite{Moffat:2021dji}, demonstrating full compatibility. 
As previously mentioned, the statistical uncertainties are currently larger than the ones from global analysis.

\section{Summary}
\label{sec:summary}

The main component of this work is the calculation of the unpolarized gluon PDF of the proton using numerical simulations of QCD. 
The calculation is performed using an $N_f=2+1+1$ ensemble of clover-improved twisted mass fermions with the quark masses tuned to give a pion mass of 260 MeV. 
The lattice spacing is 0.093 fm, and the volume is $32^3\times64$. 
For the calculation, we employ the pseudo-distribution approach that significantly simplifies the renormalization procedure by forming ratios of matrix elements, leading to the reduced pseudo-Ioffe time distributions, expressed in terms of the combination $\nu=z\cdot P$. 
In our calculation, we use a nucleon momentum boost with values up to 1.67 GeV, and, in the final results we restrict the length of the Wilson line to 0.56 fm.
We find that the combination of $P$ and $z$ suffices to extract a continuous dependence on $\nu$ and reconstruct the gluon PDF. 
We explore systematic effects such as excited-states effects, the effect of stout smearing, and the dependence on the maximum value of $z$ entering the fits to obtain the ITD. 
For the evolution and conversion to the $\MSb$ scheme at a scale of 2 GeV, we use a one-loop formalism. 
We use the fitting reconstruction method to address the inverse problem and obtain the $x$-dependence of the gluon PDF. 
A novel aspect of the calculation is the elimination of the mixing with the quark-singlet unpolarized PDF, which we extract for the same ensemble. 
The effect of the mixing brings the gluon ITD to smaller values, but the effect is much smaller than the statistical uncertainties. 
However, when the precision stage is reached for such lattice calculations, mixing will inevitably become a more important effect.
Our results are compared with other lattice data obtained using a different lattice formulation, methodology, and setup~\cite{HadStruc:2021wmh}, and we find a very good agreement.
In such a comparison, we ignore the quark-gluon mixing for a more appropriate comparison with Ref.~\cite{HadStruc:2021wmh}.
Furthermore, a comparison of our final data with the global analysis of the JAM collaboration~\cite{Moffat:2021dji} reveals agreement, with the global analysis being much more accurate than lattice data at this stage.
The above-mentioned comparison uses our data after the elimination of the quark-gluon mixing as done in JAM20.
An extension of this work is the investigation of other sources of systematic uncertainties, such as volume and discretization effects, as well as pion mass dependence. 
In the near future, we will address the continuum limit by adding two ensembles with smaller lattice spacing.

\vspace*{0.5cm}
\centerline{\textbf{Acknowledgements}}
\vspace*{0.15cm}

J. D. and M.~C. acknowledge financial support from the U.S. Department of Energy, Office of Nuclear Physics, Early Career Award under Grant No.\ DE-SC0020405.
C.A. acknowledges financial support from the project EXCELLENCE/0421/0043 "3D-Nucleon," co-financed by the European Regional Development Fund and the Republic of Cyprus through the Research and Innovation Foundation, the EU project STIMULATE that received funding from the European Union's Horizon 2020 research and innovation program under grant agreement No. 76504, and the AQTIVATE project that receives funding from the European Union’s HORIZON MSCA Doctoral Networks programme, under Grant Agreement No. 101072344.
K.~C.\ is supported by the National Science Centre (Poland) grants SONATA BIS no.\ 2016/22/E/ST2/00013 and OPUS no.\ 2021/43/B/ST2/00497.  K.H. is financially supported by the Cyprus
Research and Innovation Foundation under contract number CULTURE-AWARD-YR/0220/0012. 
Computations for this work were carried out in part on facilities of the USQCD Collaboration, which are funded by the Office of Science of the U.S. Department of Energy. It also includes calculations carried out on the HPC resources of Temple University, supported in part by the National Science Foundation through major research instrumentation grant number 1625061 and by the US Army Research Laboratory under contract number W911NF-16-2-0189.

\bibliographystyle{h-physrev}

\bibliography{references.bib}

\end{document}